\documentclass[11pt]{article}
\usepackage{caption}
\usepackage{scalerel}
\usepackage{amsthm,latexsym,amsxtra,graphicx,appendix,epstopdf,feynmf,hyperref,setspace,fix-cm}
\usepackage[normalem]{ulem}
\usepackage{makeidx}
\usepackage{fullpage}
\usepackage{amsmath}
\usepackage{amssymb}
\usepackage{setspace}
\usepackage{bbm}
\usepackage{dsfont}
\usepackage{epsfig}
\usepackage[font=footnotesize,labelfont=bf,justification=centerlast,width=.94\textwidth]{caption}
\usepackage{cite}
 \usepackage{multirow}
\usepackage{array,booktabs}
\usepackage{subfigure}
\usepackage{mathtools}
\usepackage{color}
\usepackage[makeroom]{cancel}
\usepackage{tikz}
\usepackage{pgfplots}
\usetikzlibrary{intersections, pgfplots.fillbetween}
\usepackage{tensor}
\usepackage{simplewick}
\usepackage{frcursive}

\usepackage{pgfornament,multicol}

\usepackage{caption}
\usepackage{hyperref}

\usepackage[skip=0.2cm, indent=0pt]{parskip}

\hypersetup{
    bookmarks=true,%
    colorlinks,%
    citecolor=blue,%
    filecolor=blue,%
    linkcolor=blue,%
    urlcolor=blue
}
\usepackage{float}
\usepackage{slashed}
\usepackage{tikz}
\usetikzlibrary{decorations.pathmorphing}

\newcommand{\bi}{\begin{itemize}}
\newcommand{\ei}{\end{itemize}}
\newcommand{\bea}{\begin{eqnarray}}
\newcommand{\eea}{\end{eqnarray}}
\newcommand{\be}{\begin{equation}}
\newcommand{\ee}{\end{equation}}

\def\XXint#1#2#3{{\setbox0=\hbox{$#1{#2#3}{\int}$}
     \vcenter{\hbox{$#2#3$}}\kern-.5\wd0}}

\def\={\, = \,}

\numberwithin{equation}{section}
\allowdisplaybreaks  

\begin{document}

\vspace*{2.5cm}
\begin{center}
{ \Large \textsc{Gravitational Observatories}} \\ \vspace*{2.3cm}

\end{center}

\begin{center}
Dionysios Anninos,$^{1,2}$ Dami\'an A. Galante,$^1$ and Chawakorn Maneerat$^1$  \\ 
\end{center}
\begin{center}
{
\footnotesize
\vspace{0.4cm}
$^1$Department of Mathematics, King's College London, Strand, London WC2R 2LS, UK
\\
{$^2$ Instituut voor Theoretische Fysica, KU Leuven, Celestijnenlaan 200D, B-3001 Leuven, Belgium} 
}
\end{center}
\begin{center}
{\textsf{\footnotesize{
dionysios.anninos@kcl.ac.uk, damian.galante@kcl.ac.uk, chawakorn.maneerat@kcl.ac.uk}} } 
\end{center}

\vspace*{0.5cm}

\vspace*{1.5cm}
\begin{abstract}
\noindent
We consider four-dimensional general relativity with vanishing cosmological constant defined on a manifold with a boundary. In Lorentzian signature, the timelike boundary is of the form $\boldsymbol{\sigma} \times \mathbb{R}$, with $\boldsymbol{\sigma}$ a spatial two-manifold that we take to be either flat or $S^2$. In Euclidean signature we take the boundary to be $S^2\times S^1$. We consider conformal boundary conditions, whereby the conformal class of the induced metric and trace $K$ of the extrinsic curvature are fixed at the timelike boundary. The problem of linearised gravity is analysed using the Kodama-Ishibashi formalism. It is shown that for a round metric on $S^2$ with constant $K$, there are modes that grow exponentially in time. We discuss a method to control  the growing modes by varying $K$.  The growing modes are absent for a conformally flat induced metric on the timelike boundary. We provide evidence that the Dirichlet problem for a spherical boundary does not suffer from non-uniqueness issues at the linearised level. We consider the extension of black hole thermodynamics to the case of conformal boundary conditions, and show that the form of the Bekenstein-Hawking entropy is retained.

\end{abstract}

\newpage

\setcounter{tocdepth}{2}
\tableofcontents

\section{Introduction}
\label{section: Introduction}

Depending on the physical setting, the asymptotic structure of spacetime can vary considerably. In certain circumstances, one is afforded a spacetime that asymptotes to spatial or null infinity where the dynamical behaviour of the metric can be significantly tamed. A prime example is an asymptotically anti-de Sitter spacetime whereby the metric field admits a well-posed Dirichlet condition \cite{fefferman1985conformal,KICHENASSAMY2004268} at the conformal boundary, accompanied by standard Cauchy data across a complete spacelike slice. This sets the gravitational stage \cite{deHaro:2000vlm,Skenderis:2002wp,Anderson:2004yi} for the AdS/CFT correspondence, and provides a rigid boundary coordinate system with respect to which one can characterise the dynamical features of the physical processes in the interior spacetime. The null boundary of an asymptotically Minkowski spacetime, though less rigid than that of anti-de Sitter space, is sufficiently structured to permit the gravitational scattering of asymptotic  states.\footnote{In four spacetime dimensions, the gravitational $S$-matrix suffers from infrared issues \cite{Weinberg:1965nx,Prabhu:2022zcr}, but is in the least sensible order by order in a perturbative expansion.}

There are, however, circumstances where such asymptotic structures are not available. For instance, as is often the case in cosmological models, the Cauchy surface might be a compact space such as an $S^3$. Alternatively, for spacetimes expanding at a sufficiently rapid rate, an observer may be surrounded by a cosmological horizon and hence out of causal contact from any asymptotic regions even if they were available in the global spacetime. This is of particular relevance to an asymptotically de Sitter space in which observers are surrounded by a cosmological event horizon. In such a circumstance, it is natural to ask whether one can construct a more quasi-local framework. As an auxiliary step in this direction one can imagine creating a quasi-artificial, finite-size, timelike boundary in spacetime. We can view this as a type of thickened wordline $\mathcal{M}$ \cite{Anninos:2011af,Coleman:2021nor,Banihashemi:2022jys,Witten:2023qsv,Blacker:2023oan,Loganayagam:2023pfb} in the midst of spacetime whose boundary data comprise a type of reference system. {Relatedly, in the stretched horizon picture one  imagines a timelike surface that approaches the horizon of a black hole \cite{Damour:1978cg,znajek1978electric,tHooft:1984kcu,Price:1986yy,Susskind:1993if,Bredberg:2011xw, Freidel:2023bnj} or a cosmological horizon \cite{Banks:2003cg,Banks:2006rx,Anninos:2011zn,Susskind:2021omt,Anninos:2021ihe,Shaghoulian:2021cef,Shaghoulian:2022fop,Banihashemi:2022htw}.} Notably, a worldtube  perspective appears as an important ingredient in recent literature incorporating methods of AdS/CFT to analyse the de Sitter static patch in two \cite{Anninos:2017hhn,Anninos:2018svg,Svesko:2022txo,Anninos:2022hqo} and three \cite{Coleman:2021nor,Shyam:2021ciy} dimensional models (see \cite{Galante:2023uyf} for a recent review). Moreover, the presence of a worldline plays an important role in the definition of a sensible von Neumann entropy for quantum fields weakly coupled to gravity in the static patch \cite{Chandrasekaran:2022cip}.

Given  a gravitational worldtube $\mathcal{M}$, a mathematical problem of interest is  one of well-posedness for the data on the timelike boundary $\Gamma$ of $\mathcal{M}$, where the spatial section of $\Gamma$ is taken to have finite size. In the general relativity literature, this is often referred to as an initial boundary value problem \cite{Sarbach:2012pr}. This is to be contrasted with the standard initial value problem of general relativity long established to be well-posed \cite{foures1952theoreme,Choquet-Bruhat:1969ywq}. To first approximation, one might imagine a standard Dirichlet-type boundary value problem whereby one specifies the induced metric $g_{m n}$ on $\Gamma$,  along with Cauchy data on a spacelike slice $\Sigma$ that intersects $\Gamma$. This problem has been explored for both Euclidean \cite{Anderson:2006lqb,Witten:2018lgb} and Lorentzian \cite{Friedrich:1998xt,Fournodavlos:2020wde,Fournodavlos:2021eye,An:2021fcq,Anninos:2022ujl} signature. Somewhat remarkably, and in contrast to the Klein-Gordon and Yang-Mills equations, the second order nature of the Einstein constraint equation significantly restricts the type of boundary data one can place on $\Gamma$. In particular, generic real valued Dirichlet data  do not satisfy the Einstein constraint equations at $\Gamma$ in either signature \cite{Anderson:2006lqb,An:2021fcq}. To make the discussion concrete, take the Lorentzian four-dimensional Einstein constraint equation projected along $\Gamma$ 
\begin{equation}
	\mathcal{R} - K^2 + K_{m n}K^{m n} \, |_\Gamma \= 0 \, ,
\end{equation}
where $\mathcal{R}$  denotes the Ricci scalar with respect to the metric $g_{m n}$ induced at $\Gamma$, and $K_{mn}$ denotes the second fundamental form or extrinsic curvature at $\Gamma$, with trace $K = g^{mn}K_{mn}$. If we imagine boundary data $g_{m n}$ that solves the above constraint, and vary slightly away by $\delta g_{m n}$, to linear order in $\delta g_{m n}$ one must satisfy
\begin{equation}\label{ec}
	\delta \mathcal{R} - 2 K \delta K + 2 K_{m n} \delta K^{m n} + 2 K^{m}{}_{r} K^{r n}\delta g_{m n}  \, |_\Gamma \= 0 \, .
\end{equation}
Upon taking into account tangential diffeomorphisms at $\Gamma$, the space of deformations $\delta g_{m n}$ at $\Gamma$ is captured by three independent functions. Therefore, the constraint (\ref{ec}) might only be satisfied for a subset of the full space of $\delta g_{m n}$. This reasoning has been proven for Euclidean signature \cite{Anderson:2006lqb}, as well as for a standard Minkowski corner with vanishing unperturbed $K_{mn}$ \cite{An:2021fcq}.  (In three spacetime dimensions, one has a single independent $\delta g_{m n}$ so the argument does not apply.) Similarly, one can argue that the Neumann boundary condition, whereby one instead fixes the second fundamental form $K_{m n}$ along $\Gamma$, does not have good existence properties. The presence of a non-vanishing cosmological constant $\Lambda$ does not affect the preceding reasoning, provided the boundary lies away from asymptotic boundaries. Further to the existence issues raised above, at least for certain choices of Dirichlet data on $\Gamma$ there is also a question of (non)-uniqueness that must be addressed. At least for flat boundaries there is an infinite class of physical diffeomorphisms that render the Dirichlet problem non-unique in Euclidean \cite{Anderson:2006lqb,Witten:2018lgb} and Lorentzian \cite{An:2021fcq,Anninos:2022ujl} signature.

It is useful to  contrast the generic absence of solutions to the Einstein constraint equation for a finite size boundary with the situation in a four-dimensional asymptotically AdS spacetime. There, the induced metric at the AdS boundary permits a finite neighbourhood of deformations near some reference value \cite{fefferman1985conformal,KICHENASSAMY2004268,Friedrich:1995vb}.  The reason one can more easily satisfy the constraint equation (\ref{ec}) is that near the AdS boundary one has the behaviour
\begin{equation} 
\frac{ds^2}{\ell^2} = d\rho^2 + \left( e^{2\rho} g^{(0)}_{m n} (x^m) + g^{(2)}_{m n}(x^m) +  e^{-\rho} g^{(3)}_{m n}(x^m) + \ldots \right) dx^m dx^n~,
\end{equation}
with $\rho \to \infty$, and $m$, $n$ range over the boundary coordinates $x^m$. As such, near the boundary $K_{m n} \approx e^{2\rho}  g^{(0)}_{m n} \ell$ is  parameterically large. To leading order at large $\rho$,  the cosmological extension of (\ref{ec}) is automatically satisfied for arbitrary $g_{m n}$. This can be viewed as a variant of the umbilic boundary conditions which are shown to be well-posed in \cite{Fournodavlos:2020wde, Fournodavlos:2021eye}. One can then continue solving the constraint order by order in a small $e^{-\rho}$ expansion which converges \cite{fefferman1985conformal,KICHENASSAMY2004268}. From the perspective of AdS/CFT, $g_{m n}$ constitutes a source for the dual CFT stress tensor, and is crucial to define boundary stress tensor correlation functions. More generally, from a CFT perspective one often considers coupling the CFT to a curved metric $g_{m n}$ (ideally of non-negative curvature) thereby turning on a small but finite source for the stress tensor. In Euclidean signature, the bulk solution with  prescribed boundary metric $g_{m n}$ is a real valued saddle of the Einstein equations with $\Lambda<0$. For instance, when $g_{m n}$ is the round $S^3$, we can fill the space with the Euclidean AdS$_4$ spacetime of curvature $-12/\ell^2$ and metric 
\begin{equation}
\frac{ds^2}{\ell^2} = d\rho^2 + \sinh^2 \rho \, d\Omega_3^2~,
\end{equation}
with $\rho \ge 0$. One can subsequently consider small positive curvature deviations away from the round $S^3$ and find real valued bulk solutions filling the interior. Similarly, in Lorentzian AdS one considers small but finite deformations of the boundary metric \cite{Friedrich:1995vb} leading to novel solutions in the interior. 

Perhaps all this is an indication that for boundaries of finite extent (particularly in the Euclidean case) one should consider complexified solutions, as often happens in a saddle point analysis. However, before delving into the complex plane, it is important to note that there is a set of better behaved `conformal' boundary conditions on a finite size $\Gamma$. In Euclidean signature  \cite{Anderson:2006lqb}, it has been shown that fixing the conformal class $\left[g_{m n}|_{\Gamma} \right]$ of the induced metric at the boundary and the trace $K$ of the second fundamental form $K_{m n}$ at $\Gamma$ leads to an elliptic problem (see also \cite{Figueras:2011va}). In Lorentzian signature,  it is conjectured \cite{An:2021fcq} that the same boundary conditions accompanied by standard Cauchy data on $\Sigma$ lead to a well-posed hyperbolic problem. The Lorentzian conformal boundary conditions have been shown to be well posed at the linearised level \cite{Anninos:2022ujl}. Moreover, for timelike surfaces parameterically near a black hole \cite{Bredberg:2011xw}  or cosmological  \cite{Anninos:2011zn} event horizon the conformal boundary conditions permit a rich solution space governed by a variant of the non-relativistic incompressible Navier-Stokes equation. 

This paper explores the conformal boundary conditions of \cite{Anderson:2006lqb,An:2021fcq} at the level of linearised general relativity in four spacetime dimensions with $\Lambda=0$. We consider Lorentzian boundaries of the type $\Gamma = \boldsymbol{\sigma} \times \mathbb{R}$ with $\boldsymbol{\sigma}$ a  spatial two-manifold and $\mathbb{R}$ denoting the time direction. As boundary data we append $\boldsymbol{\sigma}$ with an induced metric that is either flat or positively curved and consider a variety of extrinsic data $K$. The linearised problem is analysed through the  Kodama-Ishibashi formalism \cite{Kodama:2000fa,Kodama:2003jz}, adapted to manifolds with a boundary. The general setup is discussed in section \ref{sec:framework}. Our linearisation procedure is described in section \ref{lineargr}. Spatial boundaries of vanishing curvature with vanishing $K$ are analysed in section \ref{flatwall}, where it is shown that the perturbations are linearly stable. Spatial boundaries appended with  round two-sphere induced metric of size $\mathfrak{r}$ and $K = 2 \, \mathfrak{r}^{-1}$ are considered in \ref{sec:spherical}. It is shown that there exist linear perturbations that grow exponentially for generic initial Cauchy data, which is the conformal boundary condition counterpart of the instabilities discussed in \cite{Andrade:2015gja} for the Dirichlet problem. It is also shown that the non-uniqueness issues that appear in the flat case for the Dirichlet boundary conditions \cite{An:2021fcq,Anninos:2022ujl}  are alleviated for a spherical spatial boundary. In section \ref{sec: euclidean bh} we consider general relativity in Euclidean signature on a manifold with $S^2\times S^1$ boundary subject to conformal boundary conditions and generalise the considerations of York \cite{York:1986it} for the thermodynamics of a Schwarzschild black hole. We show that while entropy of the black hole retains the Bekenstein-Hawking form $\mathcal{S}_\text{BH} = A/4G_N$,  the specific heat and energy are modified as compared to the Dirichlet problem. Various technical details are provided in the appendices.

\section{General framework} \label{sec:framework}

\tikzset{every picture/.style={line width=0.75pt}} 

We consider vacuum solutions to general relativity in four spacetime dimensions and zero cosmological constant. In Lorentzian signature its action $I$ is given by
\begin{equation}\label{eqn: action}
	I \= \frac{1}{16 \pi G_N}\int_{\mathcal{M}} d^4 x \sqrt{-g}\, R + I_B \, ,
\end{equation}
where $G_N$ is the Newton's constant and $I_B$ is a boundary term that we will shortly specify and is needed for the variational principle to be well-defined. The equations of motion are the Einstein field equation
\begin{equation}
	R_{\mu \nu}-\frac{1}{2} g_{\mu \nu}R \= 0 \,, \label{ee}
\end{equation}
regardless of the choice of the boundary term. In what follows, we will use Greek indices, $\mu, \nu = 0,...,d$, for $(d+1)$-dimensional spacetime indices.

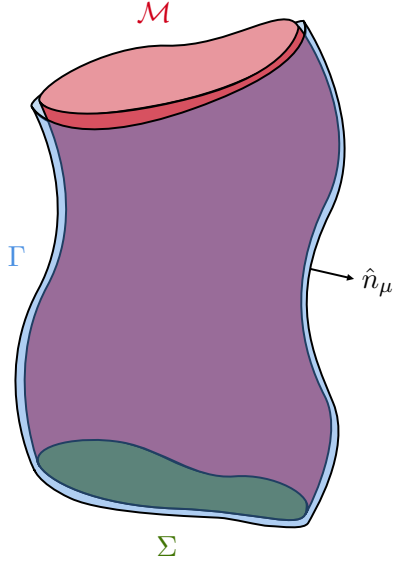
\begin{figure}[t]
	\centering

\tikzset{every picture/.style={line width=0.75pt}} 

\begin{tikzpicture}[x=0.75pt,y=0.75pt,yscale=-1,xscale=1]

\draw  [fill={rgb, 255:red, 208; green, 2; blue, 27 }  ,fill opacity=0.46 ] (271.2,260.8) .. controls (267.34,252.12) and (285.93,247.16) .. (291.75,248.25) .. controls (297.56,249.35) and (308.47,244.98) .. (335.93,258.25) .. controls (363.38,271.53) and (365.38,264.07) .. (383.38,267.53) .. controls (401.38,270.98) and (408.35,275.48) .. (406.4,285.2) .. controls (404.45,294.93) and (382.45,285.17) .. (340.95,283.67) .. controls (299.45,282.17) and (275.06,269.48) .. (271.2,260.8) -- cycle ;
\draw  [fill={rgb, 255:red, 65; green, 117; blue, 5 }  ,fill opacity=0.75 ] (271.2,260.8) .. controls (267.34,252.12) and (285.93,247.16) .. (291.75,248.25) .. controls (297.56,249.35) and (308.47,244.98) .. (335.93,258.25) .. controls (363.38,271.53) and (365.38,264.07) .. (383.38,267.53) .. controls (401.38,270.98) and (408.35,275.48) .. (406.4,285.2) .. controls (404.45,294.93) and (382.45,285.17) .. (340.95,283.67) .. controls (299.45,282.17) and (275.06,269.48) .. (271.2,260.8) -- cycle ;
\draw  [fill={rgb, 255:red, 151; green, 190; blue, 232 }  ,fill opacity=0.57 ] (268,80.82) .. controls (265.14,76.71) and (301.14,53.63) .. (333.14,54.43) .. controls (365.14,55.23) and (402.72,35.57) .. (405.11,37.69) .. controls (407.5,39.81) and (408.06,59.34) .. (380.86,71.29) .. controls (353.66,83.23) and (343.43,85.86) .. (310.57,90.43) .. controls (277.71,95) and (270.86,84.93) .. (268,80.82) -- cycle ;
\draw  [fill={rgb, 255:red, 232; green, 151; blue, 158 }  ,fill opacity=1 ] (272.4,78.8) .. controls (266.43,72.83) and (304,49.2) .. (336,50) .. controls (368,50.8) and (395.71,28.07) .. (402.4,40.4) .. controls (409.09,52.73) and (374.53,72.13) .. (331.42,80.13) .. controls (288.31,88.13) and (278.37,84.77) .. (272.4,78.8) -- cycle ;
\draw  [fill={rgb, 255:red, 215; green, 81; blue, 96 }  ,fill opacity=1 ] (402.6,40.6) .. controls (402.86,39.21) and (437,87.8) .. (416.6,128.6) .. controls (396.2,169.4) and (403.8,203.8) .. (415.8,227) .. controls (427.8,250.2) and (402.29,293.71) .. (406.6,285.4) .. controls (410.91,277.09) and (390.2,265.8) .. (370.2,267.4) .. controls (350.2,269) and (334.2,253.8) .. (313.4,250.2) .. controls (292.6,246.6) and (267.72,251.39) .. (271.4,261) .. controls (275.08,270.61) and (251,217.4) .. (275.8,171.4) .. controls (300.6,125.4) and (268.67,74.27) .. (272.6,79) .. controls (276.53,83.73) and (299.06,89.27) .. (349.4,76.2) .. controls (399.74,63.13) and (402.34,41.99) .. (402.6,40.6) -- cycle ;
\draw  [fill={rgb, 255:red, 74; green, 144; blue, 226 }  ,fill opacity=0.44 ] (405.11,37.69) .. controls (403.75,35.06) and (440.89,89.36) .. (418.67,133.8) .. controls (396.44,178.24) and (410.13,206.33) .. (420,230.24) .. controls (429.87,254.16) and (409.56,286.82) .. (407.56,290.69) .. controls (405.56,294.56) and (394,292.69) .. (388,292.47) .. controls (382,292.24) and (370,288.69) .. (362,288.24) .. controls (354,287.8) and (304.89,286.47) .. (287.56,278.47) .. controls (270.22,270.47) and (268.85,262.69) .. (269.33,264.24) .. controls (269.82,265.8) and (246.4,219.22) .. (271.2,173.22) .. controls (296,127.22) and (266.29,77.64) .. (268,80.82) .. controls (269.71,84) and (286.76,104.18) .. (349.56,81.8) .. controls (412.35,59.42) and (406.47,40.32) .. (405.11,37.69) -- cycle ;
\draw    (408.57,162.7) -- (427.65,167.08) ;
\draw [shift={(430.57,167.75)}, rotate = 192.93] [fill={rgb, 255:red, 0; green, 0; blue, 0 }  ][line width=0.08]  [draw opacity=0] (3.57,-1.72) -- (0,0) -- (3.57,1.72) -- cycle    ;

\draw (254.5,149.9) node [anchor=north west][inner sep=0.75pt]  [color={rgb, 255:red, 74; green, 144; blue, 226 }  ,opacity=1 ]  {$\Gamma $};
\draw (329.5,295.9) node [anchor=north west][inner sep=0.75pt]  [color={rgb, 255:red, 65; green, 117; blue, 5 }  ,opacity=1 ]  {$\Sigma $};
\draw (319.5,26.4) node [anchor=north west][inner sep=0.75pt]  [color={rgb, 255:red, 208; green, 2; blue, 27 }  ,opacity=1 ]  {$\mathcal{M}$};
\draw (433,160.6) node [anchor=north west][inner sep=0.75pt]    {$\hat{n}_\mu$};

\end{tikzpicture}

	\caption{Illustration of a four-manifold $\mathcal{M}$ with a timelike boundary $\Gamma$, a Cauchy surface $\Sigma$, and a unit normal vector field $\hat{n}_\mu$.}
	\label{tube}
\end{figure}

We are interested in finding solutions to (\ref{ee}), for a theory placed on a spacetime manifold $\mathcal{M}$ with a non-empty boundary $\partial \mathcal{M} \neq \emptyset$. The boundary is further composed of a spacelike boundary, that we will denote by $\Sigma$, and a timelike boundary, $\Gamma$. 
Latin indices $m,n,...$ denote  spacetime indices tangent to the timelike boundary. The general setup is depicted in figure \ref{tube}.

Following \cite{An:2021fcq}, we consider conformal boundary conditions on $\Gamma$. Concretely, we fix the conformal class of the induced metric and the trace of the extrinsic curvature on the timelike boundary
\begin{eqnarray}\label{eqn: conformal bdry data}
	\text{Conformal boundary conditions} &:&  \{ \left[g_{mn}|_{\Gamma} \right]_\text{conf}\, , \, K|_{\Gamma}  \} \quad = \quad \text{fixed} \,.
\end{eqnarray}
The above  boundary data is said to be geometric in the sense that it is constructed from quantities that are  defined on $\Gamma$ alone. The trace of the extrinsic curvature $K$ is defined as,
\begin{equation}\label{eqn: trace K}
	K \= g^{mn} K_{mn} \quad , \quad K_{mn} \= \frac{1}{2}\mathcal{L}_{\hat{n}} g_{mn} \, ,
\end{equation}
where $\hat{n} \= \hat{n}^\mu \partial_\mu$ is a unit normal vector associated to the boundary and pointing outward, and $\mathcal{L}_{\hat{n}}$ denotes a Lie derivative with respect to $\hat{n}^\mu$. We can write $K$ as a normal derivative of the boundary local volume, $K= \mathcal{L}_{\hat{n}} \log \sqrt{-\text{det}\, g_{mn} }$. The normal-normal and normal-tangential components of the Einstein field equation \eqref{ee}  projected onto $\Gamma$ are given by
\begin{equation}\label{eqn: constraints from einstein}
	\mathcal{R} - K^2 + K_{mn}K^{mn} \, |_\Gamma \= 0 \, , \qquad \qquad \mathcal{D}_m K^m{}_n - \mathcal{D}_n K \, |_\Gamma \= 0 \, ,
\end{equation}
where $\mathcal{R}$ and $\mathcal{D}_m$ denote the Ricci scalar and covariant derivative with respect to the induced metric $g_{mn}$. 
 
We emphasise that the projection of the Einstein equation on $\Gamma$, given in \eqref{eqn: constraints from einstein}, act as constraints for the boundary data of the gravitational dynamics. Note they are a non-linear combination of the induced metric $g_{mn}|_\Gamma$ and the extrinsic curvature $K_{mn}|_\Gamma$. As discussed in the introduction, specifying generic boundary data on $\Gamma$ will not satisfy these equations, causing the existence problem for the initial boundary value problem in general relativity \cite{An:2021fcq}.
 
Given the boundary conditions \eqref{eqn: conformal bdry data}, the boundary action is given by,
\begin{equation}\label{IB}
	I_B  = \frac{1}{24 \pi G_N} \int_\Gamma d^3 x \sqrt{-\text{det}\,g_{mn}}\, K \, ,
\end{equation}
which is a third of the standard Gibbons-Hawking-York term \cite{yorkbdy, Gibbons:1976ue} used for the Dirichlet problem. 
\section{Linearised gravity}\label{lineargr}

In this section, we consider perturbation about empty Minkowski spacetime, namely,
\begin{equation}
	g_{\mu \nu} \= \eta_{\mu \nu} + \epsilon \, h_{\mu \nu}  \, ,\qquad | \,\epsilon \,| \ll 1 \, .
\end{equation}
It will be convenient to write the background flat metric in Gaussian normal coordinates, that are adapted to the timelike boundary
\begin{equation}\label{eqn: choice of coordinate}
	\eta_{\mu \nu} dx^\mu dx^\nu \= dx_\perp^2 + \bar{g}_{mn} dx^m dx^n \, ,
\end{equation}
where the location of the timelike boundary $\Gamma$ is taken to be at a constant $x_\perp$, and the unit normal vector is $\hat{n} \=\partial_\perp$. As detailed in appendix \ref{app:normal}, many expressions simplify by using this metric. From now onwards, we will use this background metric to raise and lower indices. The corresponding boundary data is given by the conformal class of the three-dimensional metric $\left[g_{mn}|_{\Gamma} \right]_\text{conf}$ and the trace of the extrinsic curvature at the boundary $K|_{\Gamma}$.

\subsection{Linearised  boundary conditions}
Demanding that the conformal boundary data remains invariant under any perturbation implies that
\begin{eqnarray}
	h_{mn}|_\Gamma &=& \gamma(x) \bar{g}_{mn} |_\Gamma \,, \label{trace}\\
\epsilon \, \delta K(h_{\mu\nu})  & \equiv &	\left[ K(\eta_{\mu\nu}+\epsilon \, h_{\mu\nu})-K(\eta_{\mu\nu}) \right] |_{\Gamma}   = 0 \, ,
\end{eqnarray}
where $\gamma(x)$ is an arbitrary function, that will depend on the initial data of the linearised metric $h_{\mu \nu}$. By contracting (\ref{trace}) with $\bar{g}^{mn}$, one may rewrite the first condition as
\begin{equation}\label{eqn: bdry cond 1}
	h_{m n} - \frac{1}{3} \bar{g}_{m n } h^p{}_p  \, |_{\Gamma}  \= 0 \, .
\end{equation}
It is more convenient to use this expression instead of (\ref{trace}), as in this one, $\gamma(x)$ does not appear explicitly. Using \eqref{eqn: trace K}, it can be shown that the variation of the trace of the extrinsic curvature at a linearised level becomes
\begin{equation}
	2\delta K (h_{\mu\nu}) \= \partial_\perp h^m{}_m - 2 \mathcal{D}^m h_{\perp m} - Kh_{\perp \perp}  \, |_{\Gamma} = 0  \, ,\label{eqn: bdry cond 2}
\end{equation}
where $\mathcal{D}_n$ denotes the covariant derivative with respect to the boundary metric $\bar{g}_{mn}$. (Further details can be found in appendix \ref{app:normal}.) In our analysis we will take \eqref{eqn: bdry cond 1} and \eqref{eqn: bdry cond 2} as the conformal boundary conditions for linearised gravity.
 
We must also  declare boundary conditions for the space of allowed diffeomorphisms. At the linearised level, they act on the coordinates and metric respectively as
\begin{equation}
	x^\mu \rightarrow x^\mu -\epsilon \, \xi^\mu \, , \qquad\qquad
	h_{\mu \nu} \rightarrow h_{\mu \nu} + \nabla_\mu \xi_\nu + \nabla_\nu \xi_\mu \,,
\end{equation}
for any smooth vector field $\xi_\mu$. In the presence of a timelike boundary, one should require that diffeomorphisms leave the boundary data unchanged. This leads to boundary conditions for $\xi_\mu$, that are given by
\begin{eqnarray}\label{eqn: diffeo bdry cond 1}
	\left. \left(K_{mn} - \frac{K}{3}\bar{g}_{mn}\right)\xi_{\perp}  + \left(\mathcal{D}_{(m} \xi_{n)} - \frac{\bar{g}_{mn}}{3} \mathcal{D}^p \xi_p \right)\, \right|_{\Gamma} &=& 0 \,,  \\
	\left. \left(\partial_\perp K - \mathcal{D}^m \mathcal{D}_m \right)\xi_\perp + \xi_m \mathcal{D}^m K \, \right|_{\Gamma} &=& 0 \ \, .\label{eqn: diffeo bdry cond 2} 
\end{eqnarray}
Moreover, since we picked coordinates so that the timelike boundary $\Gamma$ is located at constant $x_\perp$, we need an additional restriction so that $\xi^\mu$ does not move $\Gamma$. Then, we will further impose that
\begin{equation}
	\xi^\perp |_\Gamma \= 0 \, . \label{eqn: diffeo bdry cond perp}
\end{equation}
The full set of boundary conditions on allowed diffeomorphisms is then given by \eqref{eqn: diffeo bdry cond 1}, \eqref{eqn: diffeo bdry cond 2}, and \eqref{eqn: diffeo bdry cond perp}. For generic background spacetimes $g_{\mu \nu}$, these reduce to setting $\xi_\mu |_\Gamma = 0$. For special sets of boundary data, the boundary conditions can allow certain non-trivial diffeomorphisms to act non-trivially on the boundary. These are precisely a set of conformal Killing vectors of the timelike boundary $\Gamma$ preserving the trace of the extrinsic curvature $K$. 
 
So far, we computed conformal boundary conditions on the timelike boundary for linearised gravity. To study different geometric configurations, it will be convenient to work in the harmonic gauge, which we discuss next. {In section \ref{sec:spherical} we will consider the gauge-invariant Kodama-Ishibashi formalism \cite{Kodama:2000fa,Kodama:2003jz}.}

\subsection{Harmonic gauge}

The advantage of working in the harmonic (or de Donder) gauge is that the linearised Einstein equation become wave equations, 
\begin{equation}\label{eqn: harmonic EOM}
-\nabla^\rho \nabla_\rho h_{\mu \nu}  \=0 \, .
\end{equation}
In the case of a non-vanishing cosmological constant, there will be an extra mass term proportional to it. At the linearised level, the harmonic gauge condition is given by imposing
\begin{equation}\label{eqn: harmonic gauge cond}
	T_\nu(h_{\mu\nu}) \,\equiv\,  \nabla^\mu h_{\mu \nu} - \frac{1}{2}\nabla_\nu h^\mu{}_\mu  \= 0 \, .
\end{equation}
These gauge conditions\footnote{For an arbitrary gravitational perturbation $\hat{h}_{\mu\nu}$, not necessarily in the harmonic gauge, the conditions \eqref{eqn: harmonic gauge cond} can be obtained by making a suitable gauge transformation,
\begin{equation}
	\hat{h}_{\mu \nu}  \rightarrow \hat{h}_{\mu \nu} + \nabla_\mu \hat{\xi}_\nu + \nabla_\nu \hat{\xi}_\mu = h_{\mu \nu} \, ,
\end{equation}
with the vector field $\hat{\xi}_\mu$ solving the inhomogeneous wave equation
\begin{equation}
	\nabla^\rho \nabla_\rho \hat{\xi}_\mu = - T_\mu (\hat{h}_{\mu\nu}) \, ,
\end{equation}
and obeying the boundary conditions \eqref{eqn: diffeo bdry cond 1}, \eqref{eqn: diffeo bdry cond 2}, and \eqref{eqn: diffeo bdry cond perp}.
}  
can be expressed in components tangent and normal to the timelike boundary as
\begin{eqnarray}\label{eqn: harmonic gauge cond 1}
	T_n(h_{\mu\nu}) & \equiv & \mathcal{D}^m h_{mn} - \frac{1}{2} \mathcal{D}_nh^m{}_m + \left(K+ \partial_\perp \right)h_{\perp n}-\frac{1}{2}\mathcal{D}_n h_{\perp \perp} = 0 \, ,\\ 
	T_\perp(h_{\mu\nu}) & \equiv & -K^{mn}h_{mn} - \frac{1}{2}\partial_{\perp} h^m{}_m + \mathcal{D}^m h_{\perp m} + K h_{\perp \perp} + \frac{1}{2}\partial_\perp h_{\perp\perp} = 0 \, . \label{eqn: harmonic gauge cond 2}
\end{eqnarray}
Since the gauge conditions must be satisfied at every point in the manifold $\mathcal{M}$ including the boundary $\partial \mathcal{M} = \Gamma \cup \Sigma$, the above equations can be regarded as providing additional boundary conditions on the metric perturbation,   $T_n(h_{\mu\nu}) |_{\Gamma} = T_\perp (h_{\mu\nu}) |_{\Gamma}=0$. So, in total, we obtain a set of four boundary conditions for the conformal boundary problem in the harmonic gauge. These are given explicitly by (\ref{eqn: bdry cond 1}) and (\ref{eqn: bdry cond 2}), as well as the two gauge conditions \eqref{eqn: harmonic gauge cond 1} and \eqref{eqn: harmonic gauge cond 2}. All in all, we have the following boundary conditions
\begin{equation}
\begin{cases}
h_{mn}-\frac{1}{3} \bar{g}_{mn}h^p{}_p \, |_{\Gamma} \= 0 \, ,\\ 
	\frac{2}{3}K h^m{}_m - \left(K+\partial_\perp\right)h_{\perp \perp} \, |_{\Gamma} \= 0 \,, \\
 	-\frac{1}{6}\mathcal{D}_n h^m{}_m + \left(K+\partial_\perp \right)h_{\perp n}-\frac{1}{2}\mathcal{D}_n h_{\perp \perp} |_{\Gamma} \=  0 \,, \\
	-\left(\frac{1}{3}K + \frac{1}{2} \partial_\perp \right) h^m{}_m + \mathcal{D}^m h_{\perp m}  + \left( K  + \frac{1}{2}\partial_\perp \right) h_{\perp \perp}  |_{\Gamma} \=0 \,.
\end{cases}
\label{eqn: harmonic bdry cond 3}
\end{equation}
We will now proceed to analyse properties of linearised gravity for a variety of boundaries subject to the above conditions.

\section{Flat boundaries}\label{flatwall}

In this section, we study the linearised gravitational problem which preserves conformal boundary data on a flat timelike boundary. More precisely, we set the timelike boundary to be locally three-dimensional Minkowski spacetime with $K=0$ and then study the behaviour of gravitational fluctuations which become conformally flat and have $\delta K(h_{\mu\nu})=0$ at the boundary. We work in the harmonic gauge (\ref{eqn: harmonic gauge cond}) in this section.
 
We consider two types of flat boundaries: a non-compact and a compact version. The former problem is equivalent to studying gravitational waves on the half-plane with conformal boundary conditions; 
in the latter case, spacetime is confined to a finite box in all spatial dimensions.\footnote{One could also consider a similar treatment for a toroidal boundary. However,  at a technical level the simplest such setup would modify the asymptotic structure of spatial infinity.}

\subsection{Non-compact}

For flat boundaries, it is convenient to use Cartesian coordinates to describe the background metric,
\begin{equation}
	ds^2 \= -dt^2 + dx^2 +dy^2 + dz^2 \, , \qquad x\geq 0 \,, \qquad t,y,z \in \mathbb{R} \,.
\end{equation}
In this setup, we consider $x_\perp = -x$, the timelike boundary $\Gamma$ to be located at $x\= 0$, and $\bar{g}_{mn}=\eta_{mn}$,. The unit normal vector associated to the timelike boundary is $\hat{n} \= -\partial_x$, namely it is pointing in an outward direction. 
 
The induced metric and the extrinsic curvature at the timelike boundary are given by
\begin{equation}\label{eqn: flat wall}
	ds^2 |_{x=0} \= -dt^2  +dy^2 + dz^2 \, , \qquad K_{mn} \= 0 \, .
\end{equation}
In the harmonic gauge, the linearised Einstein field equation is simply given by
\begin{equation}\label{eqn: eom harmonic}
	\left(-\partial_t^2 + \partial_x^2 + \partial_y^2 + \partial_z^2 \right) h_{\mu \nu}(t,\vec{x})  \= 0 \, ,
\end{equation}
where $\vec{x} \equiv (x,y,z)$. To solve this, we use the following ansatz,
\begin{equation}
	h_{\mu \nu}(t,\vec{x}) \= f_{\mu \nu}(x) e^{i k_m x^m} \, , \qquad i k_m x^m =  -i\omega t + i k_y y + i k_z z \, , \qquad k_m k^m = -k_x^2 \leq 0 \,,
\end{equation}
where we separated the $x$-dependence of $h_{\mu\nu}(t, \vec{x})$ into the function $f_{\mu \nu}(x)$. Substituting this ansatz back into (\ref{eqn: eom harmonic}), and solving the differential equation for $f_{\mu\nu}(x)$, we obtain a general metric perturbation for a given value of $\omega$, $k_y$, and $k_z$,
\begin{equation}\label{eqn: flat non-compact sol 1}
	h_{\mu \nu}(t,\vec{x}) \= 
	\begin{cases}
	 \alpha_{\mu \nu} \cos(k_x x) e^{i k_m x^m} + \beta_{\mu \nu} \sin(k_x x) e^{i k_m x^m} \, & ,  \quad  \text{if} \, \, k_nk^n \= - k_x^2 \neq 0 \,,  \\
	 \tilde{\alpha}_{\mu \nu} \, e^{i k_m x^m}+ \tilde{\beta}_{\mu \nu}\, x \,e^{i k_m x^m}\,  \, &, \quad  \text{if} \, \, k_nk^n \= - k_x^2 = 0 \,,
	 \end{cases}
\end{equation}
where $\alpha_{\mu\nu}$, $\tilde{\alpha}_{\mu\nu}$, $\beta_{\mu\nu}$, $\tilde{\beta}_{\mu\nu}$ are $k_m$-dependent functions to be determined later, and it is understood that the real part of $h_{\mu \nu}(t,\vec{x})$ must be taken. We note that the solution in the second line exhibits a behaviour linear in $x$ that might seem problematic in the $x\to \infty$ limit. However, this will not generate physical solutions with a growing mode, as shown in  section \ref{sec:kx0}. 
 
Next, we consider the conformal boundary conditions \eqref{eqn: bdry cond 1} and \eqref{eqn: bdry cond 2} in the harmonic gauge. For a flat boundary they simply become, 
\begin{eqnarray}\label{eqn: flat non-compact bdry cond 1}
	h_{m n} - \frac{1}{3} \eta_{m n} h^p{}_p \, |_{x=0} \= 0  \,, \qquad 
	\partial_x h_{x x}\, |_{x=0} \=  0\, .
\end{eqnarray}
Notice that the above equations imply that we are specifying Dirichlet boundary conditions for the non-diagonal components of $h_{mn}$ and Neumann boundary conditions on $h_{xx}$.
 
Imposing the harmonic gauge conditions further require
\begin{eqnarray}
	T_n (h_{\mu\nu}) & = & \partial^m h_{mn} - \frac{1}{2}\partial_n h^m{}_m + \partial_x h_{x n} - \frac{1}{2}\partial_n h_{x x} \= 0 \, , \label{eqn: gauge cond harmonic 1} \\
	T_x (h_{\mu\nu})  & = & -\frac{1}{2}\partial_x h^m{}_m + \partial^m h_{x m} + \frac{1}{2}\partial_x h_{x x} \= 0 \, . \label{eqn: gauge cond harmonic 2}
\end{eqnarray}
Finally, residual diffeomorphisms are described by vector fields $\xi_\mu$, which satisfy the wave equation
\begin{equation}\label{eqn: flat non-compact diff eom}
	\left(-\partial_t^2 + \partial_x^2 + \partial_y^2 + \partial_z^2 \right)\xi_\mu \= 0 \, .
\end{equation}
Analogously to the metric perturbation, solutions to this equation become,
\begin{eqnarray} \label{eqn: flat non-compact residual gauge 1} 
\xi_{\mu}(t,\vec{x}) \= 
	\begin{cases}
	 \alpha_{\mu} \cos(k_x x) e^{i k_m x^m} + \beta_{\mu} \sin(k_x x) e^{i k_m x^m} \, & ,  \quad  \text{if} \, \, k_m k^m \= - k_x^2 \neq 0 \,,  \\
	 \tilde{\alpha}_{\mu} \, e^{i k_m x^m}+ \tilde{\beta}_{\mu}\, x \,e^{i k_m x^m}\,  \, &, \quad  \text{if} \, \, k_m k^m \= - k_x^2 = 0 \,,
	 \end{cases}
\end{eqnarray}
where now the $\alpha_\mu, \tilde{\alpha}_\mu, \beta_\mu, \tilde{\beta}_\mu$ are the $k_m$-dependent functions to be determined later.
The boundary conditions for linearised diffeomorphisms \eqref{eqn: diffeo bdry cond 1} and \eqref{eqn: diffeo bdry cond 2} together with \eqref{eqn: diffeo bdry cond perp} now become
\begin{eqnarray}\label{eqn: flat non-compact diffeo bdry cond 1}
	\frac{1}{2}\left( \partial_n \xi_m + \partial_m  \xi_n \right) - \frac{1}{3}\eta_{nm} \partial^p \xi_p |_{x=0}  =  0 \,, \qquad
	\xi^x |_{x=0} =  0 \, .
\end{eqnarray}
Collecting all of these together, we can find physical  solutions to the gravitational perturbation problem with conformal boundary conditions in the presence of flat boundaries. For this, the strategy can be summarised as follows: (i) we start with the solutions \eqref{eqn: flat non-compact sol 1} (ii) we impose boundary conditions \eqref{eqn: flat non-compact bdry cond 1}, which will determine some of the coefficients of $\alpha_{\mu\nu}$, $\beta_{\mu\nu}$ (iii) we  impose the gauge constraints \eqref{eqn: gauge cond harmonic 1}, \eqref{eqn: gauge cond harmonic 2}, to further determine the remaining functions (iv) we  use residual diffeomorphisms \eqref{eqn: flat non-compact residual gauge 1} with boundary conditions \eqref{eqn: flat non-compact diffeo bdry cond 1} to fully determine the solution. What remain are  physical gravitational degrees of freedom   preserving conformal boundary data. Consistent initial data is supposed to fix the solution uniquely. We will repeat this procedure both for the $k_x\neq0$ and the $k_x \=0$ case.

 \subsubsection{$k_x\neq0$ solutions} \label{sec:kx!=0}
Recall that the metric perturbation \eqref{eqn: flat non-compact sol 1} is given by
\begin{equation}
	h_{\mu \nu}(t,\vec{x}) \= \alpha_{\mu \nu} \cos(k_x x) e^{i k_m x^m} + \beta_{\mu \nu} \sin(k_x x) e^{i k_m x^m} \, ,
\end{equation}
where it is understood that we are taking the real part. Using the above solution and the fact that $k_x \neq 0$, the boundary conditions \eqref{eqn: flat non-compact bdry cond 1} become
\begin{eqnarray}
	\alpha_{m n} - \frac{1}{3}\eta_{mn} \alpha^{p}{}_p = 0 \,, \qquad 
	\beta_{x x} = 0\, . \label{bdy_cond222}
\end{eqnarray}
Next, we need to impose the gauge constraints. Using again that $k_n k^n = -k_x^2 \neq 0$, we can define two orthogonal three-dimensional vectors that are also orthogonal to $k^n$ and have unit norm. Let us call those vectors $u^{(1)n}$ and  $u^{(2)n}$, so that
\begin{equation}
	u^{(1)m} k_m \= u^{(2)m} k_m  \= u^{(1)m} u^{(2)}_m \= 0 \, , \qquad \qquad u^{(1)m} u^{(1)}_m \= u^{(2)m} u^{(2)}_m \= 1 \,. 
\end{equation}
Now, we can first project the gauge condition $T_m(h)=0$ onto the $k^m$, $u^{(1)m}$, and $u^{(2)m}$ directions to obtain
\begin{eqnarray}\label{eqn: dummy 1}
	ik^n k^m h_{mn} + k^n \partial_x h_{x n} + i \frac{ k_x^2}{2} \left(h^m{}_m + h_{x x}\right) = 0 \,, \qquad 
	u^{(1,2)m} \left( ik^n  h_{mn} + \partial_x h_{x m} \right)= 0 \,. 
\end{eqnarray}
The other gauge condition, $T_x (h)=0$, gets simplified by acting with $\partial_x$ and imposing the equation of motion $\partial_x^2 h_{\mu \nu} \= -k_x^2 h_{\mu \nu}$, so that
\begin{eqnarray}
	k^n \partial_x h_{n x} + \frac{ik_x^2}{2} \left(- h^m{}_m +h_{x x}\right) &=& 0 \,. \label{eqn: dummy 3}
\end{eqnarray}
Inserting \eqref{eqn: flat non-compact sol 1} into \eqref{eqn: dummy 1} and \eqref{eqn: dummy 3}, and combining with the boundary conditions \eqref{bdy_cond222}, we obtain the following set of restrictions on the coefficients $\alpha_{\mu\nu}, \beta_{\mu\nu}$,
\begin{equation} \label{eqn: gauge cond kx!=0}
\begin{dcases}
               \alpha_{mn} \= 0 \,, & \left( k^m k^n + k_x^2 \eta^{mn} \right)\beta_{mn} \= 0 \,, \\
               2ik^n \beta_{x n} - k_x \alpha_{x x} = 0  \,, &  k_x  \beta^m{}_m -2ik^n \alpha_{x n} \= 0 \,, \\
               u^{(1)m} k_x \beta_{x m}  \= 0 \,,          &   i u^{(1)m} k^n \beta_{mn} - u^{(1)m} k_x \alpha_{x m}  \= 0 \,, \\
               u^{(2)m} k_x \beta_{x m}  \= 0 \,,          &   i u^{(2)m} k^n \beta_{mn} - u^{(2)m} k_x \alpha_{x m}  \= 0 \,.
        \end{dcases}
\end{equation}
Notice that the first line implies that the linearised induced metric becomes identically zero, just as if we were imposing Dirichlet boundary conditions. This is merely a coincidence for the flat boundary case. In general, it needs not to be zero, and a generic gravitational fluctuation can lead to a non-zero Weyl factor on the boundary, see, for instance, section \ref{sec:kx0}. Moreover, while this problem is well-posed, it is known that imposing Dirichlet boundary conditions on the flat boundary does violate the geometric uniqueness of the linearised solution \cite{An:2021fcq, Anninos:2022ujl}, see also section \ref{sec:dirichlet}.
 
Finally, we can use residual diffeomorphisms to completely gauge-fix the solution. Recall from \eqref{eqn: flat non-compact residual gauge 1} that we have
\begin{equation}
	\xi_\mu \= \alpha_\mu \cos(k_x x)e^{i k_n x^n} + \beta_{\mu} \sin(k_x x)e^{i k_n x^n} \, .
\end{equation}
The boundary condition for the diffeomorphisms fixes
\begin{eqnarray}
	k_n \alpha_m + k_m \alpha_n - \frac{2}{3} \eta_{n m} k^p \alpha_p &=& 0 \,, \qquad 
	\alpha_x = 0 \, .
\end{eqnarray}
For $k_n \neq 0$ (and $k_nk^n \neq 0$), the first equation implies that $\alpha_n =0$. Hence the allowed diffeomorphisms are the ones which act trivially on the timelike boundary at $x=0$. Using these residual diffeomorphisms, the remaining components of the metric perturbation transform as,
\begin{eqnarray}
\begin{cases}
	\beta_{m n} & \rightarrow \quad \beta_{mn} + i k_m \beta_n + i k_n \beta_m \,, \\
	\beta_{x n} & \rightarrow \quad \beta_{x n}  + i k_n \beta_x \,,  \\
	\alpha_{x n} & \rightarrow \quad \alpha_{x n}  + k_x  \beta_n \,,  \\
	\alpha_{x x} & \rightarrow \quad \alpha_{x x} + 2 k_x \beta_x \, .
	\end{cases}
\end{eqnarray}
We can use the last two lines to choose $\beta_\mu$ so as to fix the gauge in which $ \alpha_{x \mu}\, \= 0$. As a result, the gauge-fixed metric perturbation can be written in a compact form as 
\begin{eqnarray}\label{eqn: gauge-fixed sol kx!=0}
	h_{\mu \nu}dx^\mu dx^\nu = \left[\beta^{(+)} \left(u^{(1)}_n u^{(1)}_m - u^{(2)}_n u^{(2)}_m\right) + 2 \beta^{(\times)} u^{(1)}_n u^{(2)}_m \right]  \sin(k_x x) e^{i k_n x^n} dx^n dx^m \,,
\end{eqnarray}
where $\beta^{(+)}$ and $\beta^{(\times)}$ are built from the remaining $\beta_{mn}$ and represent the two polarisations of the gravitational fluctuation that preserve the conformal boundary data at $x=0$. We provide more details in appendix \ref{app:derivation}. Note that both gravitational modes $\beta^{(+)}$ and $\beta^{(\times)}$ are odd under $x\rightarrow -x$. Since there is no notion of a region outside the boundary with $x<0$, this property should be thought of as a mathematical feature rather than a physical symmetry of the solution.
 
\noindent \textbf{Example.} Consider the solution with $k_y\=k_z\=0$, or equivalently,
\begin{equation}
	k^n \partial_n \= \omega \partial_t \, ,
\end{equation}
where $\omega = |k_x|$. The vectors $u^{(1)n}$ and $u^{(2)n}$ can be chosen as
\begin{equation}
	u^{(1)n} \partial_n \= \partial_y \, ,\qquad \qquad u^{(2)n} \partial_n \= \partial_z \,.
\end{equation}
Then, the general metric perturbation obeying the conformal boundary conditions in this case is simply given by
\begin{equation}
	h_{\mu \nu}dx^\mu dx^\nu = \left[ \beta^{(+)}  \left( dy^2-dz^2 \right)  + 2 \beta^{(\times)} dydz\right] \sin(k_x x) e^{i k_n x^n}\, .
\end{equation}

 \subsubsection{$k_x=0$ solutions}\label{sec:kx0}
Now consider the second solution in \eqref{eqn: flat non-compact sol 1}, for which $k_x = 0$,
\begin{equation}
	h_{\mu \nu}(t,\vec{x}) \= \tilde{\alpha}_{\mu \nu} e^{i k_m x^m}+ \tilde{\beta}_{\mu \nu}\, x \,e^{i k_m x^m}\, . \label{kx0}
\end{equation}
In this case, the vector $k_n$ becomes null with respect to the boundary metric. Let $q^n$ and $u^n$ be a null and a spacelike vector, respectively, which satisfy the following properties,
\begin{equation}
	q^n k_n \= 1 \, , \qquad \qquad q^n u_n  \= k^n u_n  \= 0 \,.
\end{equation}
It is convenient to project the gauge condition $T_\mu (h_{\mu\nu})=0$ onto $k^n$, $q^n$, $u^n$ and the normal vector. Inserting the solution \eqref{kx0}, the gauge conditions become,
\begin{equation}
\begin{dcases}
             i k^m k^n \tilde{\alpha}_{mn} + k^n \tilde{\beta}_{x n} = 0\, , & k^m k^n \tilde{\beta}_{mn} \= 0 \,,  \\
	- i u^m u^n \tilde{\alpha}_{mn} + 2 q^n \tilde{\beta}_{x n} - i\alpha_{x x} = 0\,, &  2u^m u^n \tilde{\beta}_{mn}-\tilde{\beta}_{x x} \=0 \,, \\
	i k^m u^n \tilde{\alpha}_{mn} + u^n \tilde{\beta}_{x n} = 0 \,, &  k^m u^n \tilde{\beta}_{mn}\=0 \,, \\
	k^m \tilde{\alpha}_{x m} + \frac{i}{2} \left(\tilde{\beta}^m{}_m - \tilde{\beta}_{x x}\right) = 0 \,, &  k^n \tilde{\beta}_{x n}\=0 \,.
        \end{dcases}
\end{equation}
The conformal boundary conditions take a similar form to the ones in the previous subsection,
\begin{eqnarray}
	\tilde{\alpha}_{m n} - \frac{1}{3}\eta_{mn} \tilde{\alpha}^p{}_p \= 0 \, , \qquad
	\tilde{\beta}_{xx} \= 0 \, .
\end{eqnarray}
To solve the remaining gauge condition, let us look at the residual diffeomorphisms in \eqref{eqn: flat non-compact residual gauge 1},
\begin{equation}
	\xi_\mu \= \tilde{\alpha}_\mu e^{i k_n x^n} + \tilde{\beta}_{\mu} \, x \, e^{i k_n x^n} \, .
\end{equation}
Similar to the $k_x\neq 0$ case, the boundary conditions for the diffeomorphisms impose that
\begin{equation}
	k_m \tilde{\alpha}_n + k_n \tilde{\alpha}_m -\frac{2}{3}\eta_{nm} k^p \tilde{\alpha}_p \= 0 \, , \qquad \tilde{\alpha}_x \= 0 \,,
\end{equation}
which lead to $\tilde{\alpha}_\mu \= 0$. The transformation of the metric perturbation under the allowed diffeomorphisms is given by
\begin{eqnarray}
\begin{cases}
	\tilde{\alpha}_{x n} &\rightarrow \quad \tilde{\alpha}_{x n}+ \tilde{\beta}_n  \,, \nonumber\\
	\tilde{\alpha}_{x x} &\rightarrow \quad \tilde{\alpha}_{x x}+2 \tilde{\beta}_x \,, \nonumber\\
	\tilde{\beta}_{m n} &\rightarrow \quad \tilde{\beta}_{m n}+i k_m \tilde{\beta}_n + i k_n \tilde{\beta}_m  \,, \nonumber\\
	\tilde{\beta}_{x n} &\rightarrow \quad \tilde{\beta}_{x n}+  i k_n \tilde{\beta}_x \, .
	\end{cases}
\end{eqnarray}
We can use the residual diffeomorphisms to choose $\tilde{\beta}_\mu$ such that $q^m \tilde{\beta}_{m \mu} \= 0$. As a result, the general gauge-fixed metric perturbation becomes
\begin{equation}
	h_{\mu \nu}dx^\mu dx^\nu \= \tilde{\alpha}^{(+)} e^{ik_nx^n} \left(\eta_{nm}dx^n dx^m - dx^2\right) + 2 \left(\tilde{\alpha}^{(\times)} u_n + i\tilde{\alpha}^{(\perp)} k_n\right) e^{ik_n x^n}  dx^n dx \, ,
\end{equation}
where $\tilde{\alpha}^{(+)}$, $\tilde{\alpha}^{(\times)}$, and $\tilde{\alpha}^{(\perp)}$ are remaining gravitational degrees of freedom preserving the conformal boundary data at $x\=0$. Interestingly, the modes $\tilde{\alpha}^{(\times)}$ and $\tilde{\alpha}^{(\perp)}$ are odd under parity $x\rightarrow -x$ while $\tilde{\alpha}^{(+)}$ is even. In particular, the even mode becomes non-zero when evaluated at the boundary,
\begin{equation}
	h_{\mu \nu} dx^\mu dx^\nu |_{x=0} \= \tilde{\alpha}^{(+)}  \left(-dt^2 + dy^2 + dz^2\right) e^{ik_nx^n} \, ,
\end{equation}
serving as an example of a fluctuating Weyl factor that cannot be eliminated by a coordinate transformation.
 
The modes $\tilde{\alpha}^{(+)}$ and $\tilde{\alpha}^{(\times)}$ can be interpreted as two polarisations of the gravitational perturbation as they give rise to a non-vanishing Riemann tensor. On the other hand, the extra mode $\tilde{\alpha}^{(\perp)}$ leads to a vanishing Riemann tensor. In particular, it can be written locally as
\begin{equation}
	h_{\mu\nu} \= \partial_\mu \xi_\nu + \partial_\nu \xi_\mu \, , \qquad \qquad \xi_\mu \= \eta_{\mu x} \tilde{\alpha}^{(\perp)} e^{ik_n x^n} \, .
\end{equation}
Note, however, that $\xi_\mu$ is a diffeomorphism that preserves the conformal boundary data \eqref{eqn: diffeo bdry cond 1} and \eqref{eqn: diffeo bdry cond 2} but not the location of the timelike boundary, \eqref{eqn: diffeo bdry cond perp}. Consequently, it cannot be gauged away and should be treated as physical. 
 
The existence of the physical diffeomorphism mode $\tilde{\alpha}^{(\perp)}$ features a striking difference to the gravitational dynamics without boundary. In fact, by relaxing the harmonic gauge choice, one can make an arbitrary gauge transformation away from the timelike boundary, $x=0$, so that the physical diffeomorphism mode $\tilde{\alpha}^{(\perp)}$ is localised arbitrarily close to the boundary. Hence, one may call it a boundary mode.
 
\noindent \textbf{Example.} Consider the particular solution with $k_z=0$ and $k_y=\omega$. The vectors $k^n$, $q^n$, and $u^n$ are then given by
\begin{eqnarray}
	k^n \partial_n = \omega\left(\partial_t + \partial_y \right) \,, \qquad
	q^n \partial_n = \frac{1}{2 \omega}\left(-\partial_t + \partial_y \right) \,, \qquad
	u^n \partial_n = \partial_z \, .
\end{eqnarray}
The corresponding metric perturbation now reads
\begin{equation}
	h_{\mu \nu}dx^\mu dx^\nu \= \left[\tilde{\alpha}^{(+)}  \left(-dt^2+dy^2+dz^2 - dx^2\right) + 2\tilde{\alpha}^{(\times)}  dx dz + 2i\tilde{\alpha}^{(\perp)} \omega \left(-dt dx+dy dx\right)\right] e^{-i\omega(t-y)}\, .
\end{equation}

\subsection{Modifying $\delta K$}

In this section, we consider slightly changing the boundary conditions to include a non-vanishing (but small) variation of the trace of the extrinsic curvature at the timelike boundary. Consequently, the boundary conditions \eqref{eqn: flat non-compact bdry cond 1} now become,
\begin{eqnarray}\label{eqn: flat non-compact bdry cond deltaK}
	h_{m n} - \frac{1}{3} \eta_{m n} h^p{}_p \, |_{x=0} \= 0  \,, \qquad 
	\partial_x h_{x x}\, |_{x=0} \=  2 \, \delta K(t,y,z)\, ,
\end{eqnarray}
where $\delta K(t,y,z)$ is an external function that will act as the new conformal boundary data. We will now solve this problem using the solutions we found for the vanishing $\delta K$ case.
 
Consider a metric perturbation that consists of the sum of two terms
\begin{equation}\label{eqn: h_munu with delta K}
	h_{\mu \nu} \= \bar{h}_{\mu \nu} + \partial_\mu \xi_\nu + \partial_\nu \xi_\mu \,.
\end{equation}
The condition that $h_{\mu\nu}$ is in the harmonic gauge implies that $\bar{h}_{\mu\nu}$ satisfies \eqref{eqn: eom harmonic} and $T_\mu (\bar{h}_{\mu\nu})=0$, and $\xi_\mu$ satisfies \eqref{eqn: flat non-compact diff eom}. 
 
Next, we impose that, at the boundary, $\bar{h}_{\mu\nu}$ obeys the boundary conditions with $\delta K=0$, namely boundary conditions \eqref{eqn: flat non-compact bdry cond 1}. Instead, $\xi_\mu$ will obey the following boundary conditions,
\begin{equation}\label{eqn: diffeo remove delta K}
	\xi_m |_{x=0} \= 0 \, , \qquad \qquad \xi_x (t,y,z) |_{x=0} \= v(t,y,z)\, ,
\end{equation}
where the scalar function $v(t,y,z)$ satisfies the inhomogeneous wave equation
\begin{equation}\label{eqn: flat wall inhom}
	\partial_m \partial^m v(t,y,z) \= \delta K (t,y,z) \, .
\end{equation}
From here, it becomes evident that $\bar{h}_{\mu\nu}$ is the metric perturbation with vanishing trace of the extrinsic curvature previously studied and the second term, $\partial_\mu \xi_\nu + \partial_\nu \xi_\mu$, is responsible for the effect of adding $\delta K$.
 
The fact that equation \eqref{eqn: flat wall inhom} is a wave equation means that a unique $v$ can be obtained once initial data $\{v|_{t=0}, \partial_t v|_{t=0}\}$ is specified. Given such $v$, the problem of finding $\xi_\mu$ reduces to the problem of solving the wave equation of a four-dimensional vector field \eqref{eqn: flat non-compact diff eom} with Dirichlet boundary conditions \eqref{eqn: diffeo remove delta K} which is known to be well-posed.
 
Note also that this $\xi_\mu$ does not obey the boundary conditions for allowed diffeomorphisms \eqref{eqn: diffeo bdry cond perp}, so it cannot be removed by residual diffeomorphisms and consequently, it is physical.
 
We can also describe the effect of adding $\delta K$ in terms of a change in the location of the timelike boundary. Given \eqref{eqn: h_munu with delta K}, consider the following infinitesimal coordinate transformation,
\begin{equation}
	x^\mu \rightarrow x'^\mu \= x^\mu + \epsilon \, \xi^\mu(x^\mu) \, ,
\end{equation}
so that the term $\partial_\mu \xi_\nu + \partial_\nu \xi_\mu$ is removed from the metric perturbation \eqref{eqn: h_munu with delta K}. In the new coordinates $x'^\mu$, the location of the timelike boundary $\Gamma$ is mapped to $x'^\mu \= \xi^\mu|_{x=0}$ or, equivalently, 
\begin{equation}
	x' \= \epsilon \, \xi^x|_{x=0} \= \epsilon \, v(t,y,z) \, .
\end{equation}
Therefore, the general metric perturbation preserving boundary data $\delta K$ and conformally flat can be described by
\begin{equation}
	h_{\mu \nu} \= \bar{h}_{\mu\nu} \, ,
\end{equation}
on a Minkowski spacetime restricted to $ x \geq \epsilon \, v(t,y,z)$. 
 
In summary, adding a small variation $\delta K$ to the flat boundary problem amounts to imposing the dynamics \eqref{eqn: flat wall inhom} to the two-dimensional boundary of the spacelike boundary $\partial\Sigma$, which we now call a corner. To fully obtain the location of the corner, one must specify the initial location $v|_{t=0}$ and the initial velocity $\partial_tv|_{t=0}$.
 
\noindent \textbf{Example.} Finally, let us provide one simple example. Consider perturbing the trace of the extrinsic curvature by some constant $\delta K$. One solution to \eqref{eqn: flat wall inhom} is given by
\begin{equation}
	v \= -\delta K \frac{t^2}{2} \, ,
\end{equation}
which, to the linearised order, implies that the timelike boundary $\Gamma$ is moving with a constant acceleration set by $\epsilon \, \delta K$. Working in the harmonic gauge, one can extend $\xi_\mu$ into the bulk spacetime using \eqref{eqn: flat non-compact diff eom} as
\begin{equation}
	\xi_\mu dx^\mu \= -\delta K \left(\frac{t^2-x^2}{2}\right) dx \, .
\end{equation}
 
Another solution to \eqref{eqn: flat wall inhom}, with different initial data, is given by
\begin{equation}
	v \= \frac{\delta K}{2}\frac{y^2+z^2}{2} \, .
\end{equation}
This solution can be interpreted as describing the spacetime region near a timelike boundary which, at a constant time, is a two-dimensional sphere of radius $\mathfrak{r} = \frac{2}{\epsilon \, \delta K}$. Likewise, $\xi_\mu$ can be extended into the bulk as
\begin{equation}
	\xi_\mu dx^\mu \= \frac{\delta K}{2} \left(\frac{y^2+z^2}{2}-x^2\right)dx \, .
\end{equation}
 
We will study linearised solutions with spherical  boundaries with non-vanishing trace of the extrinsic curvature in more detail in section \ref{sec:spherical}.

\subsection{Dirichlet flat boundary} \label{sec:dirichlet}
 
Before continuing, it is instructive to briefly summarise the equivalent problem but with Dirichlet boundary conditions, as studied in \cite{Anninos:2022ujl}. 
  
If instead of imposing conformal boundary conditions, we impose Dirichlet boundary conditions, we would need to satisfy
\begin{equation}
	h_{mn} |_{x=0} \= 0 \,.
\end{equation}
In the $k_x\neq0$, this is equivalent to setting $\alpha_{mn} \= 0$ in \eqref{eqn: flat non-compact sol 1}. As a consequence, the harmonic gauge conditions become 
\begin{eqnarray}
\begin{dcases}
	-2ik^n \beta_{x n} +k_x \alpha_{x x}  = 0 \,, & 2ik^n \alpha_{x n} + k_x \beta_{x x} - k_x \eta^{mn} \beta_{mn} = 0 \, , \\
	u^{(1)n} \beta_{x n}  = 0 \,, & u^{(1)n} k_x \alpha_{x n} -  i u^{(1)m} k^n \beta_{m n} = 0 \, ,\\
	u^{(2)n} \beta_{x n}  = 0 \,, & u^{(2)n} k_x \alpha_{x n} -  i u^{(2)m} k^n \beta_{m n} = 0 \, ,\\
	\left(k^m k^n + k_x^2 \eta^{m n }\right) \beta_{mn} = 0 \,,  & 
	\end{dcases}
\end{eqnarray}
where $u^{(1)m}$ and $u^{(2)m}$ are defined in section \ref{sec:kx!=0}. The allowed diffeomorphisms in this case are the same as in \eqref{eqn: flat non-compact residual gauge 1},
\begin{equation}
	\xi_\mu \= \alpha_\mu \cos(k_x x)e^{i k_n x^n} + \beta_{\mu} \sin(k_x x)e^{i k_n x^n} \, .
\end{equation}
Requiring that $\xi_\mu$ preserves the induced metric and the location of the timelike boundary leads to the following boundary conditions,
\begin{equation}
	k_m \alpha_n + k_n \alpha_m \= 0 \, , \qquad \qquad \alpha_x \= 0 \, ,
\end{equation}
which imply that $\alpha_m=0$ for $k_x \neq 0$. Using these residual diffeomorphism, the metric perturbation transforms as
\begin{eqnarray}
\begin{cases}
	\beta_{m n} & \rightarrow \quad \beta_{mn} + i k_m \beta_n + i k_n \beta_m \,, \\
	\beta_{x n} & \rightarrow \quad \beta_{x n}  + i k_n \beta_x \,,  \\
	\alpha_{x n} & \rightarrow \quad \alpha_{x n}  + k_x  \beta_n \,,  \\
	\alpha_{x x} & \rightarrow \quad \alpha_{x x} + 2 k_x \beta_x \, .
	\end{cases}
\end{eqnarray}
Using the residual diffemorphism $\beta_\mu$, we can choose the gauge fixing conditions
\begin{equation}\label{eqn: gauge choice Dirichlet}
	k^m\beta_{mn} \= 0 \, , \qquad \qquad \beta_{xx} \= i \alpha_{xx} \, . 
\end{equation}
As a result, the gauge-fixed metric perturbation is given by
\begin{eqnarray}\label{eqn: sol Dirichlet}
	h_{\mu \nu}dx^\mu dx^\nu &=& \left[\beta^{(+)} \left(u^{(1)}_m u^{(1)}_n - u^{(2)}_m u^{(2)}_n\right) + 2 \beta^{(\times)} u^{(1)}_m u^{(2)}_n \right]  \sin(k_x x) e^{i k_n x^n} dx^m dx^n  \nonumber \\
	& & \quad + 2 i \alpha^{(\perp)} e^{i k_x x + i k_n x^n} \left(k_x dx^2 + k_n dx^n dx\right)\, ,
\end{eqnarray}
where $\beta^{(+)}$, $\beta^{(\times)}$, and $\alpha^{(\perp)}$ are induced metric-preserving gravitational degrees of freedom after removing gauge ambiguity. The two polarisations of the gravitational perturbation are represented by $\beta^{(+)}$ and $\beta^{(\times)}$, while $\alpha^{(\perp)}$ corresponds to a physical diffeomorphism in the same spirit as in section \ref{sec:kx0}. The main difference here is that the physical diffeomorphism mode $\alpha^{(\perp)}$ for Dirichlet problem exists even when $k_x \neq 0$. 

\noindent \textbf{Geometric uniqueness.} The existence of multiple solutions satisfying the same Dirichlet boundary conditions can be seen as follows. We can choose initial data to set $\beta^{(+)} = \beta^{(\times)} = 0$, so that the metric perturbation is governed by the $\alpha^{(\perp)}$ mode. This mode takes the form of a right-moving wave in the $x$-direction for $k_x \geq 0$. Following the argument in \cite{Anninos:2022ujl}, it is possible to construct waves which are initially localised in the $(y,z)$-direction and have compact support in the negative-$x$ region. The existence of such solutions therefore spoils the geometric uniqueness of the linearised Einstein field equation with Dirichlet boundary conditions.
 
Note also that since the non-uniqueness is directly connected to the mode $\alpha^{(\perp)}$ which is a physical diffeomorphism, any results obtained from naive gauge-invariant quantities need to be handled carefully as it sees such mode as physically trivial.
 
Finally, we emphasise that although the existence of right-moving solutions might seem as an artifact of choosing the gauge \eqref{eqn: gauge choice Dirichlet}, there exist gauge-invariant arguments of non-uniqueness of this Dirichlet problem (in Euclidean signature) \cite{Witten:2018lgb}.

\subsection{Compact boundary}

In this section, we consider a similar conformal boundary problem but in a finite spatial region. Namely, we consider spacetime inside a timelike box, whose boundary is composed of six flat timelike boundaries. 
 
Starting with the standard Minkowski metric,
\begin{equation}
	ds^2 \= -dt^2 + dx^2 + dy^2 + dz^2 \, ,
\end{equation}
the manifold $\mathcal{M}$ of interest is a region restricted to $x \in [0,L_x]$, $y \in [0,L_y]$, and $z \in [0,L_z]$ while $t \in \mathbb{R}$. Its timelike boundary is given by $\Gamma = \partial \mathcal{M} = \bigcup_{i=1}^6 \Gamma_i $, where $\Gamma_i$ is a previously studied flat timelike boundary located at $x=0$, $x=L_x$, $y=0$, $y=L_y$, $z=0$, or $z=L_z$. It follows that each side of the box $\Gamma_i$ has the same boundary data, {i.e.} conformally flat and vanishing $K$. For simplicity, we will set the length of all sides to be equal, $L_x=L_y=L_z=L$. 
 
Working in the harmonic gauge, gravitational dynamics in the interior of the box are similarly governed by the wave equation \eqref{eqn: eom harmonic} and the gauge constraints \eqref{eqn: gauge cond harmonic 1} and \eqref{eqn: gauge cond harmonic 2}, as in the non-compact version. The difference here is that now, the boundary conditions \eqref{eqn: flat non-compact bdry cond 1} have to be imposed on all the six sides of the box.  
 
 Let us first consider the case when the momentum of the gravitational perturbation is non-zero in all components. Let $n_{(i)}$ be the normal vector of the side $\Gamma_i$. We consider gravitational fluctuation with the momentum $k^\mu n_{(i)\mu} \neq 0$ for $i=1,2,...,6$. 
  
 Near each timelike boundaries $\Gamma_i$, the metric perturbation behaves according to the result found in section \ref{sec:kx!=0}. This suggests that we should look for a solution which behaves like either $\sin$ or $\cos$ functions in all $x,y,z$ coordinates.
  
 For example, the boundary conditions \eqref{eqn: flat non-compact bdry cond 1} combining with the gauge constraints, \eqref{eqn: gauge cond harmonic 1} and \eqref{eqn: gauge cond harmonic 2}, lead to
 \begin{equation}
 	h_{\mu \nu}\left(\delta^\mu_\rho- n^\mu_{(i)} n_{(i)\rho}\right)\left(\delta^\nu_\sigma- n^\nu_{(i)} n_{(i)\sigma}\right)\Big|_{\Gamma_i} \= 0 \, , \qquad \qquad n^\rho_{(i)} \partial_\rho \left(h_{\mu \nu}n^{\mu}_{(i)} n^{\nu}_{(i)} \right)\Big|_{\Gamma_i} \= 0 \, ,
 \end{equation}
 for $i=1,2,...,6$. The first condition means that the metric perturbation projected onto the timelike boundary $\Gamma_i$ obeys the Dirichlet boundary condition. Similarly, the second condition imposes the Neumann boundary condition on the normal-normal component of the metric perturbation. Boundary conditions on the remaining components then follow from the gauge constraints $T_\mu(h_{\mu\nu})\=0$.
  
As an example, let us look at a particular component $h_{xx}$. The boundary conditions imposed on $h_{xx}$ are given by
 \begin{equation}
 	h_{xx}|_{y=0 \cup y=L \cup z=0 \cup z=L} \= 0 \, , \qquad \qquad \partial_x h_{xx}|_{x=0 \cup x=L} \= 0 \, .
 \end{equation}
 A solution to the equation \eqref{eqn: eom harmonic} subject to these boundary conditions is 
 \begin{equation}
 	h_{xx} \= \alpha_{xx} \cos\left(k_x x\right) \sin\left(k_y y\right) \sin\left(k_z z\right) e^{-i \omega t}\, , \qquad \left(k_x,k_y,k_z\right) \= \frac{ \pi}{L} \left(n_x,n_y,n_z\right) \, ,
 \end{equation}
 where $n_x,n_y,n_z = 1,2,3,...$ are non-zero integers, $\omega = \sqrt{  k_x^2 + k_y^2 +k_z^2 }$, and $\alpha_{xx}$ is a $k_\mu$-dependent function. Note that, as expected for the dynamics of waves in a confined region, the gravitational fluctuation in the conformal box is characterised by a discrete set of frequencies $\omega$. 
  
 To obtain the gauge-fixed metric perturbation, we consider residual diffeomorphisms $\xi_\mu$. The harmonic gauge condition require $\xi_\mu$ to satisfy the wave equation \eqref{eqn: flat non-compact diff eom}. In section \ref{sec:kx!=0}, we showed that the boundary conditions \eqref{eqn: flat non-compact diffeo bdry cond 1} in the harmonic gauge simplify to
 \begin{equation}
 	\xi_\mu |_{\Gamma_i} \= 0 \, ,
 \end{equation}
 for $i=1,2,...,6$. Therefore, the residual diffeomorphism is given by
 \begin{equation}
 	\xi_\mu \= \beta_{\mu} \sin(k_x)\sin(k_y)\sin(k_z)e^{-i\omega t} \, , \qquad \left(k_x,k_y,k_z\right) \= \frac{ \pi}{L} \left(n_x,n_y,n_z\right) \, ,
 \end{equation}
 where $\beta_{\mu}$ are $k_\mu$-dependent functions. We then use $\beta_\mu$ to further fix some components of $h_{\mu\nu}$ to zero.
  
 For the sake of completeness, we present here the gauge-fixed gravitational perturbation subject to the conformal boundary conditions of the timelike box:
 \begin{eqnarray}
 	h_{\mu \nu} &=& h_{\mu\nu}^{(+)} + h_{\mu\nu}^{(\times)} \, , \\
	h_{\mu\nu}^{(+)}dx^\mu dx^\nu &=& \beta^{(+)}e^{-i \omega t}\Bigg( \frac{k_yk_z}{\omega^2} \cos(k_x x)\sin(k_y y)\sin(k_z z) dt dx - \frac{k_xk_y}{\omega^2} \sin(k_x x)\sin(k_y y)\cos(k_z z) dt dz  \nonumber \\
	&& + \frac{ik_z}{\omega} \cos(k_x x)\cos(k_y y)\sin(k_z z) dx dy - \frac{ik_x}{\omega} \sin(k_x x)\cos(k_y y)\cos(k_z z) dy dz \Bigg)  \, , \\
	h_{\mu\nu}^{(\times)} dx^\mu dx^\nu &=& \beta^{(\times)}e^{-i \omega t}\Bigg( \frac{k_yk_z}{\omega^2} \cos(k_x x)\sin(k_y y)\sin(k_z z) dt dx - \frac{k_xk_z}{\omega^2} \sin(k_x x)\cos(k_y y)\sin(k_z z) dt dy  \nonumber \\
	&& + \frac{ik_y}{\omega} \cos(k_x x)\sin(k_y y)\cos(k_z z) dx dz - \frac{ik_x}{\omega} \sin(k_x x)\cos(k_y y)\cos(k_z z) dy dz \Bigg)  \, ,
 \end{eqnarray}
 where 
 \begin{equation}
 	(k_x,k_y,k_z) \= \frac{\pi}{L}(n_x,n_y,n_z) \, , \qquad \omega \= \sqrt{k_x^2+k_y^2+k_z^2}\, .
 \end{equation}
 $\beta^{(+)}$ and $\beta^{(\times)}$ are $k_\mu$-dependent functions representing gravitational polarisations. No gravitational mode gives rise to a non-vanishing Weyl factor when $k^\mu n_{(i)\mu} \neq 0$ for all $i=1,2,...,6$.
  
 Now we consider the case when the momentum $k^\mu$ is pointing parallel to, at least, one side of the box, namely $k^\mu n_{(i)\mu}=0$ for some $i$. For concreteness, we set $k_x\=0$ while $k_y,k_z \neq 0$. It follows that near the boundaries $x\=0,L$, the metric perturbation now behaves as in section \ref{sec:kx0}, that is the ansatz is taken to be a function linear in $x$ and $\sin$ and $\cos$ in $y,z$.
  
 In the harmonic gauge, the residual diffeomorphism $\xi_\mu$ is a solution to \eqref{eqn: flat non-compact diff eom} subject to boundary conditions
 \begin{equation}
 	\xi_\mu |_{\Gamma_i} \= 0 \, ,
 \end{equation}
 on all sides $i=1,2,...,6$. Since the $x$-dependence of $\xi_\mu$ is given by a linear function, there is no non-trivial way to satisfy these boundary conditions at both $x=0$ and $x=L$. Hence, the only residual diffeomorphism is the trivial one $\xi_\mu \= 0$, and no further gauge fixing is needed.
  
Thus, for $k_x\=0$, a gauge-fixed metric perturbation preserving conformal boundary data of the timelike box is given by $h_{\mu \nu} = h_{\mu\nu}^{(\times)} + \partial_{\mu}\xi_{\nu}+ \partial_{\nu}\xi_{\mu}$, with
   \begin{eqnarray}
 \begin{cases}
	h_{\mu\nu}^{(\times)} dx^\mu dx^\nu = \tilde{\beta}^{(\times)}e^{-i \omega t}\Bigg( \frac{ik_y}{\omega} \sin(k_y y)\cos(k_z z) dx dz - \frac{ik_z}{\omega} \cos(k_y y)\sin(k_z z) dx dy \Bigg)  \, , \\
	\xi_\mu dx^\mu = \left(\tilde{\alpha}^{(\perp)} +\tilde{\beta}^{(\perp)} x\right) e^{-i\omega t}\sin(k_y y) \sin(k_z z) dx \, ,
	\end{cases}
 \end{eqnarray}
 where $\tilde{\beta}^{(\times)}$, $\tilde{\alpha}^{(\perp)}$, and $\tilde{\beta}^{(\perp)}$ are $k_\mu$-dependent functions. The first mode $\tilde{\beta}^{(\times)}$ represents a gravitational polarisation confined to a box, while $\tilde{\alpha}^{(\perp)}$, and $\tilde{\beta}^{(\perp)}$ are physical diffeomorphisms. 
 
\noindent \textbf{Geometric uniqueness.} Similar to the non-compact boundary setup, the same argument can be applied here to show that there exist multiple solutions exhibiting the same initial and Dirichlet boundary data. As an example, let us consider the following solution.
\begin{equation}
	h_{\mu\nu} \= \partial_\mu \xi_\nu + \partial_\nu \xi_\mu \, , \qquad \xi_\mu dx^\mu \= \alpha^{(\perp)}e^{-i\omega t + ik_x x} \sin(k_y y)\sin(k_z z) dx \, ,
\end{equation}
where $\omega \= \sqrt{k_x^2 + k_y^2 + k_z^2}$ and $\left(k_y , k_z\right) \= \frac{\pi}{L}\left(n_y,n_z\right)$. The momentum $k_x$ is not restricted to a multiple of integer. This mode takes the form of a wave moving in an $x$-increasing direction. As a result, one can construct waves which initially have support outside the box. Then, at a later time, these waves will move into the interior of the box and therefore spoil the geometric uniqueness of the linearised Einstein field equation with Dirichlet boundary conditions.
 
\section{Spherical boundaries} \label{sec:spherical}

In this section, we study the linearised gravitational problem which preserves conformal boundary data on a timelike boundary with spherical symmetry. In particular, we set the timelike boundary to be a three-dimensional timelike tube of constant radius endowed with a metric of a three-dimensional cylinder and $K \= $ constant. We then study gravitational fluctuations which leave the cylinder's conformal structure unchanged and have $\delta K \=0$.
 
More precisely, by working in standard spherical coordinates,
\begin{equation}\label{eqn: flat spherical coord}
	ds^2 \= -dt^2 + dr^2 + r^2 d\Omega^2 \,, \qquad d\Omega^2 \,\equiv\, d\theta^2 + \sin^2\theta d\phi^2 \, ,
\end{equation}
we consider $x_\perp \= r$, the timelike tube $\Gamma$ to be located at constant $r=\mathfrak{r}$, and 
\begin{equation}
	\bar{g}_{mn}dx^m dx^n \= -dt^2 + r^2 d\Omega^2 \, .
\end{equation}
We are interested in the spacetime region inside the timelike tube, namely we consider $r \in \left(0, \mathfrak{r}\right)$, $t \in \mathbb{R}$, and $\theta \in (0,\pi)$, $\phi \sim \phi+2\pi$ to be angular coordinates parameterising the unit two-sphere. 
 
The induced metric and extrinsic curvature associated to the tube are given by
\begin{equation}\label{eqn: spherical wall}
	ds^2 |_{r=\mathfrak{r}} \= -dt^2  +\mathfrak{r}^2 d\Omega^2 \, , \qquad K_{mn}dx^mdx^n \= \mathfrak{r} \, d\Omega^2 \, ,
\end{equation}
implying that $K={2}/{\mathfrak{r}}$ is a positive constant. The size of the tube $\mathfrak{r}$ is thus set by the trace of the extrinsic curvature $K$.
 
We note that there is another natural choice of spacetime corresponding to the region outside of the tube, $r \in \left(\mathfrak{r},\infty\right)$. For this, the normal vector is pointing in an opposite direction which results in a negative sign of the extrinsic curvature $K_{mn}$.\footnote{The convention we use here is that the normal vector is pointing outward with respect to the region of interest.} Although we will not consider this case here in detail, but briefly discuss it in subsection \ref{l2modes}.
 
Given the spherical symmetry, it is convenient to decompose the metric perturbation $h_{\mu\nu}$ into irreducible representations of the two-sphere. This can be done by using the spherical harmonics decomposition of a rank-2 symmetric tensor. Consequently, the general metric perturbation $h_{\mu \nu}$ is organised by angular momentum $l \in \mathbb{N}_0$.
 
For lower angular momentum, $l\=0,1$, we work in the harmonic gauge.
 
For $l \geq 2$, we use the Kodama-Ishibashi formalism, which we review in Appendix \ref{sec: Kodama-Ishibashi}. This formalism has the benefit of dealing with gauge invariant quantities.
However, as we already encountered in the case of flat boundaries (see section \ref{sec:kx0}), there can be physical gravitational fluctuations that can be described as pure gauge solutions, so we must consider them separately.
 
Regardless of the choice of gauge and angular momentum $l$, the metric perturbations are required to obey the conformal boundary conditions \eqref{eqn: bdry cond 1} and \eqref{eqn: bdry cond 2}. For gravitational perturbations inside the worldtube, we also require the perturbations to be regular at the origin. The only exception is a time-independent, spherically symmetric perturbation with $l=0$ which, as we will see, can be treated as describing a small black hole.

\subsection{$l=0$ mode}\label{sec: l=0}

In this section, we study spherically symmetric metric perturbations. The general metric perturbation preserving spherical symmetry can be written as
\begin{equation}
	h_{\mu \nu}dx^\mu dx^\nu \= h_{tt} dt^2 + h_{r r} dr^2 + 2 h_{t r} drdt + \gamma \, r^2 d\Omega^2  \, ,
\end{equation}
where $h_{tt}$, $h_{tr}$, $h_{rr}$, and $\gamma$ are functions of $t$ and $r$ only. In the harmonic gauge, the linearised Einstein field equation is given in 
\eqref{eqn: harmonic EOM}, that in this coordinate system becomes
\begin{eqnarray}
\begin{cases}
	\left(-\partial_t^2 + \partial_r^2 + \frac{2}{r} \partial_r  \right)h_{tt} &= 0 \, , \\
	\left(-\partial_t^2 + \partial_r^2 + \frac{2}{r} \partial_r - \frac{2}{r^2}\right) h_{ tr} &= 0 \, , \\
	\left(-\partial_t^2 + \partial_r^2 + \frac{2}{r} \partial_r \right) \left(h_{rr}+2\gamma\right) &= 0 \, , \\
	\left(-\partial_t^2 + \partial_r^2 + \frac{2}{r} \partial_r - \frac{6}{r^2} \right)\left(h_{rr}-\gamma \right) &= 0  \, .
	\end{cases}
\end{eqnarray}
Note that they can all be recast in the form
\begin{equation}\label{eqn: spherical Bessel diff eqn}
	\left(-\partial_t^2 + \partial_r^2 + \frac{2}{r}\partial_r - \frac{q(q+1)}{r^2}\right) F_q (t,r) \= 0 \, ,
\end{equation}
for some function $F_q (t,r)$ with $q \in \mathbb{N}_0$. To solve this equation, it is convenient to use a time Fourier transform ansatz,
\begin{equation}
	F_q(t,r) \= f_q(r) e^{-i \omega t},
\end{equation}
where the $r$-dependence is solely encoded in the unknown functions $f_q(r)$. Plugging this ansatz back into \eqref{eqn: spherical Bessel diff eqn} and solving the differential equation for $f_n(r)$, we obtain a general solution for a given $\omega$,
\begin{eqnarray} \label{eqn: spherical bessel diff sol}
F_q(\alpha, \beta; t,r) \= 
	\begin{cases}
	 \alpha \, J_q (r \omega) e^{- i \omega t} + \beta \, Y_q (r \omega) e^{-i\omega t} \, & ,  \quad  \text{if} \, \, \omega \, \neq \, 0 \,,  \\
	 \tilde{\alpha} \, r^q + \tilde{\beta} \, r^{-1-q}  \,  \, &, \quad  \text{if} \, \, \omega \=  0 \,,
	 \end{cases}
\end{eqnarray}
where $\alpha$, $\tilde{\alpha}$, $\beta$ and $\tilde{\beta}$ are $\omega$-dependent functions to be determined later. The functions $J_q (r\omega)$ and $Y_q (r\omega)$ are the spherical Bessel functions of first and second kind, respectively. 
The different metric perturbations are therefore given by 
\begin{equation}\label{eqn: spherical wall l=0 metric sol}
	h_{tt} \= F_0(\alpha_{tt},\beta_{tt}) \, , \quad h_{tr} \= F_1(\alpha_{tr},\beta_{tr}) \, , \quad h_{rr}+2\gamma \= F_0(\alpha_{+},\beta_{+}) \, , \quad h_{rr}-\gamma \= F_2(\alpha_{-},\beta_{-}) \, ,
\end{equation}
where $\alpha_{tt}$, $\alpha_{tr}$, $\alpha_{+}$, $\alpha_{-}$, $\beta_{tt}$, $\beta_{tr}$, $\beta_{+}$, $\beta_{-}$ are $\omega$-dependent functions. We omitted the $(t,r)$ dependence here for brevity. When $\omega=0$, the solutions are the same but with tilde coefficients.
 
Note that near the origin $r\rightarrow 0$, $Y_q (r \omega)$ becomes singular while $J_q (r \omega)$ remains regular for $q >0$. So,  regularity near the origin fixes all the $\beta$'s to zero.

Next, we consider the conformal boundary conditions \eqref{eqn: bdry cond 1} and \eqref{eqn: bdry cond 2}. In the harmonic gauge, they are given by 
\begin{equation}\label{eqn: spherical wall bdry cond}
	h_{tt} + \gamma \, |_{r=\mathfrak{r}} \= 0 \, , \qquad \qquad \frac{4}{\mathfrak{r}}\gamma - \left(\frac{2}{\mathfrak{r}} +\partial_r \right)h_{r r}   \, |_{r=\mathfrak{r}} \= 0\, ,
\end{equation}
where we have used the data from \eqref{eqn: spherical wall}.
 
Additionally, we need to impose the harmonic gauge conditions $T_\mu (h_{\mu\nu})\=0$. In this case, the only non-trivial components become 
\begin{eqnarray}
\begin{cases}
	T_t (h_{\mu \nu}) =& - \frac{1}{2}\partial_t \left(h_{tt}+2 \gamma \right)  + \left(\frac{2}{r}+\partial_r\right) h_{t r} - \frac{1}{2}\partial_t h_{r r} \= 0  \, , \\
	T_r (h_{\mu \nu}) =& - \frac{2}{r}\gamma + \frac{1}{2}\partial_r \left(h_{tt}-2\gamma\right) - \partial_t h_{tr} + \left(\frac{2}{r}+\frac{1}{2}\partial_r\right)h_{rr} \= 0 \, . \label{eqn: spherical wall gauge constraint}
	\end{cases}
\end{eqnarray}
 
Lastly, we consider a residual diffeomorphism $\xi_\mu$. In the spherically symmetric sector, the general diffeomorphism is given by
\begin{equation}
	\xi_\mu dx^\mu \= \xi_t dt + \xi_r dr \, ,
\end{equation}
where $\xi_t$ and $\xi_r$ are functions of $t$ and $r$ only. The harmonic gauge constraints impose that they satisfy a vector Laplacian in spherical coordinates,
\begin{eqnarray}
\begin{cases}
	\left(-\partial_t^2 + \partial_r^2 + \frac{2}{r}\partial_r \right)\xi_t &= 0 \, , \\
	\left(-\partial_t^2 + \partial_r^2 + \frac{2}{r}\partial_r - \frac{2}{r^2}\right) \xi_r &=0 \, .
\end{cases}\label{eqn: l=0 eom of diffeo}
\end{eqnarray}
These equations are of the type \eqref{eqn: spherical Bessel diff eqn}, and so the residual diffeomorphism is given by
\begin{equation}\label{eqn: sphere l=0 diffeo sol}
	\xi_t \= F_0(\alpha_t,\beta_t) \, , \qquad \xi_r \= F_1 (\alpha_r,\beta_r) \, ,
\end{equation} 
where $\alpha_t$, $\alpha_r$, $\beta_t$, and $\beta_r$ are $\omega$-dependent functions. The boundary conditions for the linearised diffeomorphism read
\begin{equation} \label{eqn: spherical diffeo bdry cond}
	\partial_t \xi_t |_{r=\mathfrak{r}} \= 0 \, , \qquad \qquad \xi^r |_{r=\mathfrak{r}} \= 0 \,,
\end{equation}
which can be derived from \eqref{eqn: diffeo bdry cond 1}, \eqref{eqn: diffeo bdry cond 2}, and \eqref{eqn: diffeo bdry cond perp}.

Under this diffeomorphism, the metric perturbation transforms as
\begin{eqnarray}
\begin{cases}
	h_{tt} & \rightarrow \quad h_{tt} + 2 \partial_t \xi_t \, , \\
	h_{tr} & \rightarrow \quad h_{tr} + \partial_t \xi_r + \partial_r \xi_t \, , \\
	h_{rr} & \rightarrow \quad h_{rr} + 2 \partial_r \xi_r \, , \\
	\gamma & \rightarrow \quad \gamma + \frac{2}{r}\xi_r \, .
	\end{cases}
\end{eqnarray}

\subsubsection{Time-dependent solutions}\label{sec: l=0 time dependent}

We first consider the time-dependent solution, namely $\omega \,\neq\, 0$. Requiring that solution is smooth everywhere in the interior restricts the solution \eqref{eqn: spherical wall l=0 metric sol} to be
\begin{eqnarray}
\begin{cases}
	h_{tt} &=\quad \alpha_{tt} J_0 (r\omega) e^{-i \omega t} \, , \\
	h_{tr} &=\quad \alpha_{tr} J_1 (r\omega) e^{-i \omega t}  \, , \\
	h_{rr} &=\quad \left(\frac{\alpha_+}{3}J_0(r\omega) + \frac{2\alpha_-}{3}J_2(r\omega)\right)e^{-i\omega t} \, , \\
	\gamma &=\quad \left(\frac{\alpha_+}{3}J_0(r\omega) - \frac{\alpha_-}{3}J_2(r\omega)\right)e^{-i \omega t} \, .
	\end{cases}
\end{eqnarray}
 
Next, we impose the harmonic gauge constraints. A direct computation shows that the harmonic constraints \eqref{eqn: spherical wall gauge constraint} yield
\begin{equation}
	\alpha_- \= - \alpha_+ \, , \qquad \qquad i\alpha_{tr} \= \frac{\alpha_+ + \alpha_{tt}}{2} \, .
\end{equation}
As a result, the general spherically symmetric time-dependent metric perturbation can be written locally as a pure gauge solution,
\begin{equation}\label{eqn: l=0 omega !=0 sol}
	h_{\mu \nu} \= \nabla_\mu \xi_\nu + \nabla_\nu \xi_\mu \, ,  \qquad \qquad \xi_\mu dx^\mu \= \left(\frac{i\alpha_{tt}}{2 \omega} J_0(r\omega) dt + \frac{\alpha_+}{2 \omega} J_1(r\omega)dr \right) e^{-i \omega t}\, .
\end{equation}
 
This result is consistent with the fact that, in the spherically symmetric sector, every smooth metric perturbation can be written locally as a pure gauge solution, a perturbative version of Birkhoff's theorem \cite{Kodama:2003jz}.  However, in the presence of the timelike boundary, this does not neccesarily mean that the gravitational dynamics is trivial, as shown in section \ref{sec:kx0} when considering a flat timelike boundary. 
 
We still need to impose the conformal boundary conditions. Inserting the solution \eqref{eqn: l=0 omega !=0 sol} into \eqref{eqn: spherical wall bdry cond}, the boundary conditions become
\begin{equation}
	\alpha_{tt} J_0(\mathfrak{r} \, \omega)+\frac{\alpha_{+}}{\mathfrak{r} \, \omega}J_1(\mathfrak{r} \, \omega) \= 0 \, , \qquad \qquad \left(2+\mathfrak{r}^2 \omega^2\right) \alpha_{+}J_1(\mathfrak{r} \, \omega) \= 0 \, .
\end{equation}
To satisfy these equations, we need a relation between $\alpha_{tt}$ and $\alpha_+$ and a condition restricting admissible values of the frequencies $\omega$. We find three inequivalent ways of satisfying those equations. Namely,
\begin{eqnarray}
\begin{cases}
	\alpha_{+} \= 0 &\text{and}\quad J_0(\mathfrak{r} \, \omega) \= 0 \, , \\
	\alpha_{tt} \= 0 &\text{and}\quad J_1(\mathfrak{r} \, \omega) \= 0 \, , \\
	\alpha_{tt} \= - \frac{ J_1(\mathfrak{r} \, \omega)}{\mathfrak{r} \, \omega J_0(\mathfrak{r} \, \omega)}\alpha_+ &\text{and}\quad \mathfrak{r}^2 \omega^2 \= - 2 \, .
	\end{cases}
\end{eqnarray}
The first and second cases fix the frequencies to be zeros of $J_0(\mathfrak{r} \, \omega)$ and $J_1(\mathfrak{r} \, \omega)$, respectively. 
In terms of $\xi_\mu$, it can be seen that they also obey the boundary conditions for the allowed diffeomorphism \eqref{eqn: spherical diffeo bdry cond}. Hence, these solutions can be gauged away using an appropriate gauge transformation. The last case restricts $\omega$ to be purely imaginary implying that the associated metric perturbation has exponential behaviour. As opposed to the first two cases, this solution does not obey the boundary conditions \eqref{eqn: spherical diffeo bdry cond} and so, it should be treated as physical. 
 
Hence, the most general metric perturbation is given by $h_{\mu \nu} = \nabla_\mu \xi_\nu + \nabla_\nu \xi_\mu$, with
\begin{eqnarray}
	\xi_\mu dx^\mu &=& \alpha^{(+)}\left(-\frac{J_1(\sqrt{2}i)}{\sqrt{2} J_0(\sqrt{2}i)} J_0(i \sqrt{2} r/\mathfrak{r})dt + J_1(i\sqrt{2}r/\mathfrak{r})dr\right)e^{\sqrt{2} t/\mathfrak{r}}  \nonumber \\
	&&+  \alpha^{(-)}\left(\frac{J_1(-\sqrt{2}i)}{\sqrt{2} J_0(-\sqrt{2}i)} J_0(-i \sqrt{2} r/\mathfrak{r})dt + J_1(-i\sqrt{2}r/\mathfrak{r})dr\right)e^{-\sqrt{2} t/\mathfrak{r}} \, , \label{eq_diff}
\end{eqnarray}
where $\alpha^{(+)}$ and $\alpha^{(-)}$ represent gravitational modes which preserve conformal structure and the trace of the extrinsic curvature at the boundary. By evaluating \eqref{eq_diff} at the boundary, we find that this solution also gives rise to a non-vanishing Weyl factor,
\begin{equation}
	h_{\mu \nu}dx^\mu dx^\nu |_{r=\mathfrak{r}} \= \left(\frac{2 \alpha^{(+)}}{\mathfrak{r}}J_1(i\sqrt{2})e^{\sqrt{2}t/\mathfrak{r}} - \frac{2 \alpha^{(-)}}{\mathfrak{r}}J_1(-i\sqrt{2})e^{-\sqrt{2}t/\mathfrak{r}} \right)\left(-dt^2 + \mathfrak{r}^2 d\Omega^2\right) \, .
\end{equation}
 
It is interesting to note that although the $\alpha^{(+)}$ mode grows exponentially, the Riemann tensor remains zero at all times since the solution can be written as a pure diffeomorphism.

\subsubsection{Time-independent solutions}

Now we consider the solution \eqref{eqn: spherical wall l=0 metric sol} with $\omega \= 0$,
\begin{eqnarray}
\begin{cases}
	h_{tt} &=\quad \tilde{\alpha}_{tt}  + \frac{\tilde{\beta}_{tt}}{r} \, , \\
	h_{tr} &=\quad \tilde{\alpha}_{tr} r + \frac{\tilde{\beta}_{tr}}{r^2}  \, , \\
	h_{rr} &=\quad \frac{\tilde{\alpha}_+}{3} + \frac{2\tilde{\alpha}_-}{3}r^2 + \frac{\tilde{\beta}_+}{3r} + \frac{2\tilde{\beta}_-}{3r^3} \, , \\
	\gamma &=\quad \frac{\tilde{\alpha}_+}{3} - \frac{\tilde{\alpha}_-}{3}r^2 + \frac{\tilde{\beta}_+}{3r} - \frac{\tilde{\beta}_-}{3r^3}  \, .
	\end{cases}
\end{eqnarray}
In this case, the harmonic gauge constraints \eqref{eqn: spherical wall gauge constraint} become algebraic equations for the eight constants of integration, $\tilde{\alpha}_{tt}$, $\tilde{\alpha}_{tr}$, $\tilde{\alpha}_{+}$, $\tilde{\alpha}_{-}$, $\tilde{\beta}_{tt}$, $\tilde{\beta}_{tr}$, $\tilde{\beta}_{+}$, and $\tilde{\beta}_{-}$. They can be solved by setting
\begin{equation}\label{eqn: sphere l=0 gauge constraints}
	\tilde{\alpha}_- \= 0 \, , \qquad \tilde{\alpha}_{tr} \= 0 \, , \qquad \tilde{\beta}_{tt} \= \frac{\tilde{\beta}_+}{3} \, .
\end{equation}
 
Next, we consider the boundary conditions \eqref{eqn: spherical wall bdry cond}, that become
\begin{equation}
	\tilde{\alpha}_{tt} + \frac{\tilde{\alpha}_+}{3} - \frac{\tilde{\alpha}_-}{3}\mathfrak{r}^2 + \frac{\tilde{\beta}_{tt}}{\mathfrak{r}} + \frac{\tilde{\beta}_+}{3\mathfrak{r}}-\frac{\tilde{\beta}_-}{3\mathfrak{r}^3} \= 0 \, , \qquad 
	\frac{2\tilde{\alpha}_+}{3\mathfrak{r}}-4\tilde{\alpha}_-\mathfrak{r} + \frac{\tilde{\beta}_+}{\mathfrak{r}^2} - \frac{2 \tilde{\beta}_-}{3\mathfrak{r}^4} \= 0 \,.
\end{equation}
Combined with the gauge constraints \eqref{eqn: sphere l=0 gauge constraints}, we find that
\begin{equation}\label{eqn: sphere l=0 bdry cond}
	\tilde{\alpha}_+ \= -\frac{3\tilde{\beta}_+}{2\mathfrak{r}}+\frac{\tilde{\beta}_-}{\mathfrak{r}^3} \, , \qquad \qquad \tilde{\alpha}_{tt} \= - \frac{\tilde{\beta}_+}{6\mathfrak{r}} \, .
\end{equation}
Thus, imposing gauge constatins \eqref{eqn: sphere l=0 gauge constraints} and boundary conditions \eqref{eqn: sphere l=0 bdry cond} reduce the number of degrees of freedom down to three.
 
Now we consider a time-independent residual diffeomorphism \eqref{eqn: sphere l=0 diffeo sol}, 
\begin{eqnarray}
	\xi_t = \tilde{\alpha}_{t}  + \frac{\tilde{\beta}_{t}}{r} \, , \qquad
	\xi_r = \tilde{\alpha}_{r} r + \frac{\tilde{\beta}_{r}}{r^2} \, .
\end{eqnarray}
The boundary conditions \eqref{eqn: diffeo bdry cond 1}, \eqref{eqn: diffeo bdry cond 2}, and \eqref{eqn: diffeo bdry cond perp} impose that 
\begin{equation}
	\tilde{\beta}_{r} \= - \tilde{\alpha}_r \mathfrak{r}^3 \, .
\end{equation}
Under this residual diffeomorphism, the metric perturbation transforms as 
\begin{eqnarray}
\begin{cases}
	\tilde{\beta}_{tr} &\rightarrow \quad \tilde{\beta}_{tr} - \tilde{\beta}_t  \, , \\
	\tilde{\alpha}_+ &\rightarrow \quad \tilde{\alpha}_+ + 6 \tilde{\alpha}_r  \, , \\
	\tilde{\beta}_{-} &\rightarrow \quad \tilde{\beta}_{-} + 6 \tilde{\alpha}_r \mathfrak{r}^3  \, .
	\end{cases}
\end{eqnarray}
Then, we can use $\tilde{\beta}_t$ and $\tilde{\alpha}_r$ to further fix the gauge $\tilde{\beta}_{tr}\=\tilde{\beta}_-\=0$. So, we are left with only one degree of freedom.
 
Note that there is a constant diffeomorphism $\tilde{\alpha}_t$ which does not change the metric perturbation, and therefore, can be omitted. In terms of the change of coordinates, it corresponds precisely to a global time translation by an infinitesimal constant $t \rightarrow t + \epsilon \, \tilde{\alpha}_t$.
 
Putting everything together, the gauge-fixed time-independent metric perturbation preserving the conformal boundary data of the timelike tube is given by
\begin{equation}\label{eqn: small bh}
	h_{\mu \nu}dx^\mu dx^\nu \= \frac{\tilde{\beta}_+}{3}\left[\left(\frac{1}{r}-\frac{1}{2\mathfrak{r}}\right)dt^2+\left(\frac{1}{r}-\frac{3}{2\mathfrak{r}}\right)\left(dr^2+r^2 d\Omega^2\right)\right]  \, .
\end{equation}
The ${1}/{r}$ behaviour suggests that the solution describes a small black hole with a singularity at $r\=0$. Its event horizon is located at a zero of $-1+ \epsilon \, h_{tt}$ which is, to first order in $\epsilon$,
\begin{equation}
	r_\text{horizon} \= \epsilon \, \frac{\tilde{\beta}_+}{3} \, .
\end{equation}
 
The appearance of the constant terms in \eqref{eqn: small bh} is crucial in maintaining the right conformal boundary data at $r=\mathfrak{r}$. In the absence of the timelike tube, one would be able to remove such terms by an appropriate rescaling of $t$ and $r$. 
 
By evaluating at $r=\mathfrak{r}$, we find that
\begin{equation}
	h_{\mu \nu}dx^\mu dx^\nu \= -\frac{\tilde{\beta}_+}{6}\left(-dt^2 + \mathfrak{r}^2 d \Omega^2\right) \, ,
\end{equation}
which implies that the size of the tube is reduced as we add a black hole in its interior. We will further explore this feature in section \ref{sec: euclidean bh} when we consider full non-linear (Euclidean) black hole solutions inside the timelike tube.

\subsection{$l=1$ mode}\label{sec: spherical l=1 mode}

Now we consider gravitational perturbations with angular momentum $l=1$. In the absence of a finite timelike boundary, any $l=1$ mode is known to be pure gauge.\footnote{To show this, start from a general $l=1$ solution to the wave equation $
	\Box h_{\mu \nu} = 0$. Then, imposing the harmonic gauge condition $T_{\mu}(h_{\mu\nu})=0$ fixes the metric perturbation to be pure gauge.}
	\, This means that the general metric perturbation in this sector can be written as $h_{\mu \nu} \= \nabla_\mu \xi_\nu + \nabla_\nu \xi_\mu \,$, with
\begin{equation}\label{eqn: l=1 ansatz}
 \xi_\mu dx^\mu \= \mathcal{T}_t \, \mathbb{S} \, dt + \mathcal{T}_r \, \mathbb{S} \, dr + \left(\mathcal{L}^S \mathbb{S}_i + \mathcal{L}^V \mathbb{V}_i \right) r \, d\Omega^i \, ,
\end{equation}
where $d \Omega^i \= (d\theta,d\phi)$. The harmonic tensors $\mathbb{S}$, $\mathbb{S}_i$, and $\mathbb{V}_i$ are  specific angular-dependent tensors that ensure that $h_{\mu\nu}$ transforms in the $l=1$ representation, see appendix \ref{sec: two-sphere}. The metric coefficients $\mathcal{T}_t$, $\mathcal{T}_r$, $\mathcal{L}^S$, and $\mathcal{L}^V$ are functions of $t$ and $r$ only.
 
The harmonic gauge constraint $T_\mu(h_{\mu\nu})\=0$ can be satisfied by requiring that $\xi_\mu$ obeys a vector Laplacian equation in spherical coordinates, leading to
\begin{eqnarray}
\begin{cases}
	\left(-\partial_t^2 + \partial_r^2 + \frac{2}{r} \partial_r - \frac{2}{r^2}  \right) \mathcal{T}_t &= 0 \, , \\
	\left(-\partial_t^2 + \partial_r^2 + \frac{2}{r} \partial_r - \frac{2}{r^2}  \right) \mathcal{L}^V &= 0 \, , \\
	\left(-\partial_t^2 + \partial_r^2 + \frac{2}{r} \partial_r \right) \left(\sqrt{2} \mathcal{T}_r - 2 \mathcal{L}^S \right) &= 0 \, , \\
	\left(-\partial_t^2 + \partial_r^2 + \frac{2}{r} \partial_r - \frac{6}{r^2} \right)\left( \sqrt{2} \mathcal{T}_r + \mathcal{L}^S \right) &= 0  \, .
	\end{cases}
\end{eqnarray}
These equations are again of the type \eqref{eqn: spherical Bessel diff eqn} and can be satisfied by the functions $F_q$ defined in \eqref{eqn: spherical bessel diff sol}. We also require that the metric perturbation with angular momentum $l=1$ is smooth everywhere in the bulk spacetime. As a result, we find that
\begin{equation}\label{eqn: spherical l=1 sol}
	\mathcal{T}_t \= F_1(\alpha_t,0) \, , \quad \mathcal{L}^V \= F_1(\alpha_v,0) \, , \quad \sqrt{2}\mathcal{T}_r - 2 \mathcal{L}^S \= F_0(\alpha_-,0) \, , \quad \sqrt{2}\mathcal{T}_r + \mathcal{L}^S \= F_2(\alpha_+,0) \, ,
\end{equation}
where $\alpha_t$, $\alpha_v$, $\alpha_-$, and $\alpha_+$ are $\omega$-dependent functions.
 
The conformal boundary conditions \eqref{eqn: bdry cond 1} and \eqref{eqn: bdry cond 2} further require that
\begin{equation}\label{eqn: spherical wall l=1 bdry cond}
	\left. \partial_t \mathcal{L}^V \right|_{r=\mathfrak{r}} \= 0 \, , \,
	\left. \partial_t^2 \mathcal{T}_r \right|_{r=\mathfrak{r}} \= 0 \, , \,
	\left. \left( r \, \partial_t \mathcal{L}^S - \sqrt{2} \mathcal{T}_t \right) \right|_{r=\mathfrak{r}} \= 0 \, , \,
	\left. \left( \sqrt{2}\mathcal{T}_r + \mathcal{L}^S + \sqrt{2} \, r\, \partial_t \mathcal{T}_t \right) \right|_{r=\mathfrak{r}} \= 0 \, .
\end{equation}
Inserting solutions \eqref{eqn: spherical l=1 sol}, these equations give relations among the different $\alpha$ functions. Solving them, we then obtain a general metric perturbation in $l=1$ sector preserving the conformal boundary data on the timelike tube at $r=\mathfrak{r}$.
 
Recall that any solution that also satisfies \eqref{eqn: diffeo bdry cond perp} can be gauged away by an allowed residual diffeomorphism. In this case, this condition becomes just $\mathcal{T}^r |_{r=\mathfrak{r}} \= 0$, 
so to obtain physical modes, it is sufficient to look only at solutions with 
\begin{equation}
	\mathcal{T}^r |_{r=\mathfrak{r}} \neq 0.
\end{equation} 

The second boundary condition in \eqref{eqn: spherical wall l=1 bdry cond} then implies that we must set $\omega \= 0$, {i.e.}, only time-independent solutions are physical.\footnote{The only exception is when $\mathcal{T}_r|_{r=\mathfrak{r}}$, $\mathcal{L}^S|_{r=\mathfrak{r}}$ are linear in $t$ and $\mathcal{T}_t|_{r=\mathfrak{r}}$ is a constant. By imposing the harmonic gauge constraint and the conformal boundary conditions, this diffeomorphism simply becomes a generator of a Lorentz boost, which leads to a vanishing metric perturbation.} This turns all the $\alpha$'s into $\tilde{\alpha}$'s. Then, the most general smooth time-independent solution \eqref{eqn: spherical l=1 sol} is given by
\begin{equation}\label{eqn: spherical l=1 time-independent sol}
	\mathcal{T}_t \= \tilde{\alpha}_t r \, , \qquad \mathcal{L}^V \= \tilde{\alpha}_v r \, , \qquad \mathcal{T}_r \= \sqrt{2} \tilde{\alpha}_+ r^2 - \frac{\tilde{\alpha}_-}{\sqrt{2}} \, , \qquad \mathcal{L}^S \= \tilde{\alpha}_+ r^2 + \tilde{\alpha}_-  \, .
\end{equation}
The boundary conditions \eqref{eqn: spherical wall l=1 bdry cond} in the time-independent case are simplified to
\begin{equation}\label{eqn: spherical wall l=1 time-independent bdry cond}
	\left. \mathcal{T}_t \right|_{r=\mathfrak{r}} \= 0 \, , \qquad \qquad \left. \sqrt{2}\mathcal{T}_r +\mathcal{L}^S \right|_{r=\mathfrak{r}} \= 0 \, .
\end{equation}
From here we obtain that $\tilde{\alpha}_t \= \tilde{\alpha}_+ \= 0$. Also, $\tilde{\alpha}_v$ can be set to zero by a residual diffeomorphism. 
 
As a result, the physical metric perturbation in $l=1$ sector obeying the conformal boundary condition is described just by the $\tilde{\alpha}_-$ mode,
\begin{equation}
	\xi_\mu dx^\mu \= -\frac{\tilde{\alpha}_-}{3} \left(\frac{\mathbb{S}}{\sqrt{2}} \, dr - \mathbb{S}_i \, r \,d\Omega^i \right) \, .
\end{equation}
However, for this solution, we find that 
$	h_{\mu \nu} \= \nabla_\mu \xi_\nu + \nabla_\nu \xi_\mu \= 0 $, so the metric perturbation identically vanishes.

\noindent \textbf{Example.} Consider the case in which $\mathbb{S}$ and $\mathbb{S}_i$ are $\phi$-independent. Their explicit expressions are given by
\begin{equation}
	\mathbb{S} \= \sqrt{\frac{3}{4 \pi}} \cos\theta \, , \qquad \mathbb{S}_\theta \= \sqrt{\frac{3}{8 \pi}} \sin\theta \, , \qquad \mathbb{S}_\phi \= 0 \, .
\end{equation}
The $\tilde{\alpha}_-$ mode now reads
\begin{equation}
	\xi_\mu dx^\mu \= -\tilde{\alpha}_- \sqrt{\frac{3}{8 \pi}}\left( \cos\theta dr - r\sin\theta d\theta \right) \= - \tilde{\alpha}_- \sqrt{\frac{3}{8 \pi}} d \left(r \cos\theta\right)\,,
\end{equation}
which corresponds to an infinitesimally global translation. Then, we can simply ignore this mode.
 \newline\newline
In summary, there is no gravitational fluctuation around flat spacetime in the $l=1$ sector which preserves the conformal boundary data on the timelike tube. The question of whether the $\tilde{\alpha}_-$ mode can become physical when a black hole or a cosmological constant is included remains an interesting open question.\footnote{We note that, in the case of Dirichlet boundary conditions investigated in \cite{Andrade:2015gja}, similar physical diffeomorphisms occur in $l=1$. In the case where there is a black hole in the interior and Dirichlet boundary conditions, the associated metric perturbation is not vanishing. The resulting linearised solutions then describe the motion of the black hole centre of mass relative to the timelike boundary.}

\subsection{$l\geq2$ modes}\label{l2modes}

Now we consider gravitational fluctuations with angular momentum $l \geq 2$. We first study gravitational perturbations in the Kodama-Ishibashi formalism. From a local point of view, these are gauge-invariant perturbations. In the presence of a boundary, one could also have physical diffeomorphisms. 

We show in a gauge-invariant way that there are no physical diffeomorphisms with angular momentum $l\geq2$  preserving the conformal boundary data of the timelike tube.
 
 \textbf{Kodama-Ishibashi formalism.}  
For a given $l \geq 2 $, the two polarisations of the gravitational fluctuations can be divided into vector and scalar 
perturbations which are encoded in master fields $\Phi^{V}(t,r)$ and $\Phi^{S}(t,r)$, respectively. As detailed in appendix \ref{sec: Kodama-Ishibashi}, let us decompose our metric perturbation into a scalar and vector perturbation as follows
\begin{equation}
    h_{\mu \nu} \= h_{\mu \nu}^{(V)} + h_{\mu \nu}^{(S)} \, ,
\end{equation}
where
\begin{eqnarray}
\begin{cases}
    h_{a b}^{(V)} &=\, 0 \, , \\
    h_{a i}^{(V)} &=\, rf^V_a \,\mathbb{V}_i \, , \\
    h_{ij}^{(V)} &=\, 2r^2H^V \,\mathbb{V}_{ij} \, ,  
    \end{cases}
    \qquad \qquad 
\begin{cases}
    h_{a b}^{(S)} &=\, f_{ab}\,\mathbb{S} \, , \\ 
    h_{a i}^{(S)} &=\, rf^S_a \,\mathbb{S}_i  \, , \\
    h_{ij}^{(S)} &=\, 2r^2 H^S \,\mathbb{S}_{ij} + 2r^2\gamma \, \sigma_{ij} \mathbb{S}\, .
 \end{cases}
\end{eqnarray}
Here $a,b = \{t,r \}$ and $i,j = \{ \theta,\phi \}$, whilst  $\mathbb{S}$, $\mathbb{S}_i$, $\mathbb{S}_{ij}$, $\mathbb{V}_i$, and $\mathbb{V}_{ij}$ are angle-dependent tensors, defined in appendix \ref{sec: Kodama-Ishibashi}, ensuring that $h_{\mu\nu}$ transforms in the $l$ representation of $SO(3)$. The coefficients of the perturbation, $f_{ab}$, $f_{a}^{V/S}$, $H^{V/S}$, and $\gamma$, are functions of the  coordinates $(t,r)$. Finally, $\sigma_{ij}$ are the metric components of the round two-sphere.

To explicitly express the metric perturbation $h_{\mu\nu}$ in terms of the master fields, it is useful to fix the following gauge:
\begin{equation}\label{eqn: sphere l>2 gauge fixing}
	H^V \= H^S \= f_t^S \=  f_{tt} + 2 \gamma \= 0  \, .
\end{equation}
Using \eqref{eqn: sphere l>2 gauge fixing}, \eqref{eqn: KI gauge invariant 1}, and \eqref{eqn: KI gauge invariant 2}, the metric perturbation is given in terms of the master fields by
\begin{eqnarray}
\begin{cases}
	h_{mn} &= - \bar{g}_{mn} \frac{1}{r}\left[\frac{l(l+1)}{2}+r^2\partial_t^2+r\partial_r\right]\Phi^S \mathbb{S} + \left(\delta_{m}^i \delta_{n}^t+\delta_{n}^i \delta_{m}^t\right) \partial_r \left(r \, \Phi^V\right)\mathbb{V}_i \, , \\
    h_{rr} &= -\frac{1}{r}\left[\frac{3l(l+1)}{2}+ 3r^2\partial_t^2 + \left(l(l+1) + 1 + r^2\partial_t^2\right) r\partial_r\right]\Phi^S \mathbb{S} \, , \\
    h_{tr} &=  -\frac{1}{2}\partial_t \left[l(l+1) - 2 + r^2\partial_t^2\right]\Phi^S \mathbb{S} \, ,  \\
    h_{ri} &=  \frac{\sqrt{l(l+1)}}{2}\left[l(l+1)+r^2\partial_t^2+2r\partial_r\right]\Phi^S \mathbb{S}_i - r \partial_t \Phi^V \mathbb{V}_i \, ,	
    \end{cases}\label{eqn: spherical l>2 ansatz}
\end{eqnarray}
where $m,n=\{t,\theta,\phi\}$.
Using the gauge-fixed metric perturbation \eqref{eqn: spherical l>2 ansatz}, the linearised Einstein field equation becomes
\begin{equation}\label{eqn: sphere master field}
	\left(- \partial_t^2 + \partial_r^2 - \frac{l(l+1)}{r^2}\right)\Phi^{V/S} \= 0 \, ,
\end{equation}
for both $\Phi^V$ and $\Phi^S$. To solve this, we use the time-Fourier transform ansatz,
\begin{equation}
	\Phi^{V/S}(t,r) \= r\omega \phi^{V/S}(r) e^{-i \omega t},
\end{equation}
where $\phi^{V/S}(r)$ are functions of $r$, and $\omega$ is an arbitrary constant. Plugging back to equation \eqref{eqn: sphere master field} and solving a differential equation of $\phi^{V/S}$, we obtain a general solution of master fields,
\begin{equation}\label{eqn: spherical l>2 sol}
	\Phi^{V/S} \= r \omega \Big( \alpha^{V/S} J_l(r\omega) + \beta^{V/S} Y_l (r\omega) \Big) e^{-i \omega t} \, ,
\end{equation}
where $\alpha^{V/S}$, $\beta^{V/S}$ depend on $\omega$ and $l$. The $J_l (r\omega)$ and $Y_l (r\omega)$ are spherical Bessel functions of the first and second kind of order $l$, respectively. Requiring that the metric perturbation is smooth everywhere in the interior sets $\beta^V \= \beta^S \= 0$. 
 
\begin{figure}[h!]
        \centering
         \subfigure[Vector, $l = 2$]{
                \includegraphics[scale=0.55]{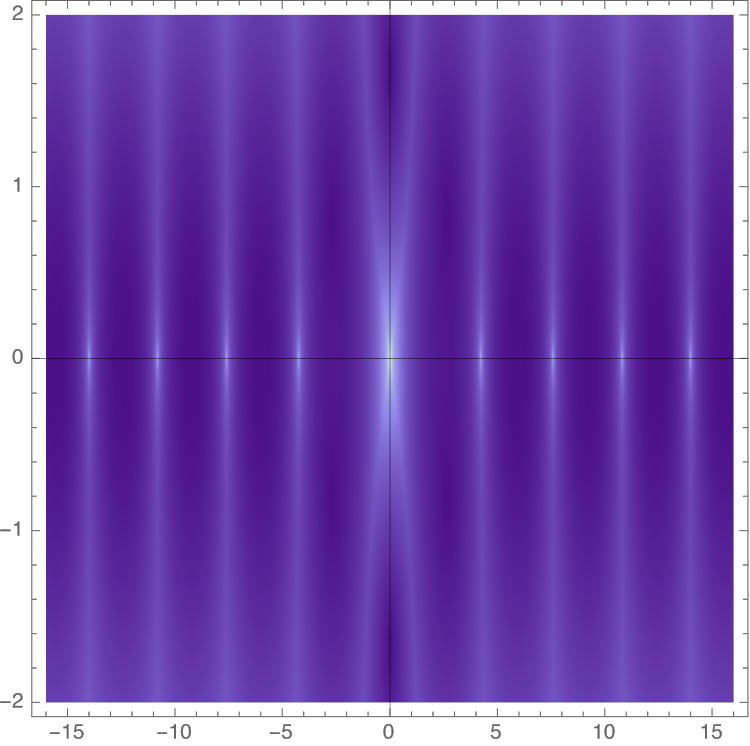}\label{q_2_4}}  \quad\quad
                 \subfigure[Vector, $l = 4$]{
                \includegraphics[scale=0.55]{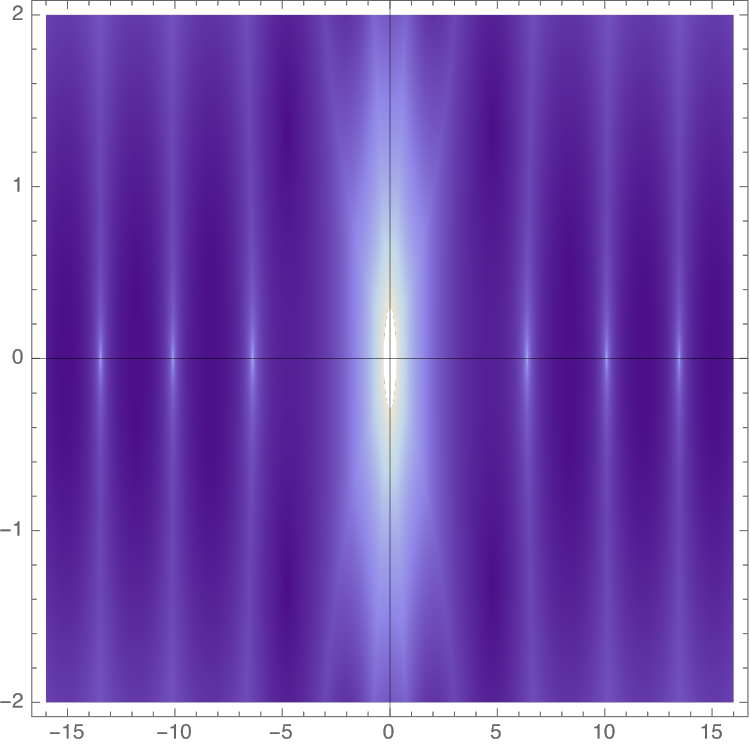} \label{q_2_4i}}  \quad\quad
        \subfigure[Scalar, $l = 2$]{
                \includegraphics[scale=0.55]{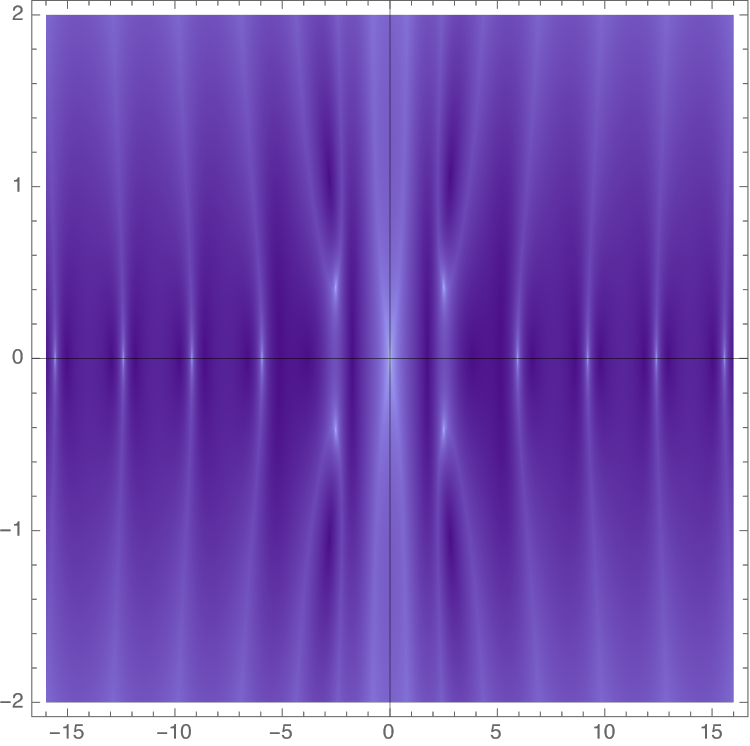} \label{q_4_8}}  \quad\quad
             \subfigure[Scalar, $l = 4$]{
                \includegraphics[scale=0.55]{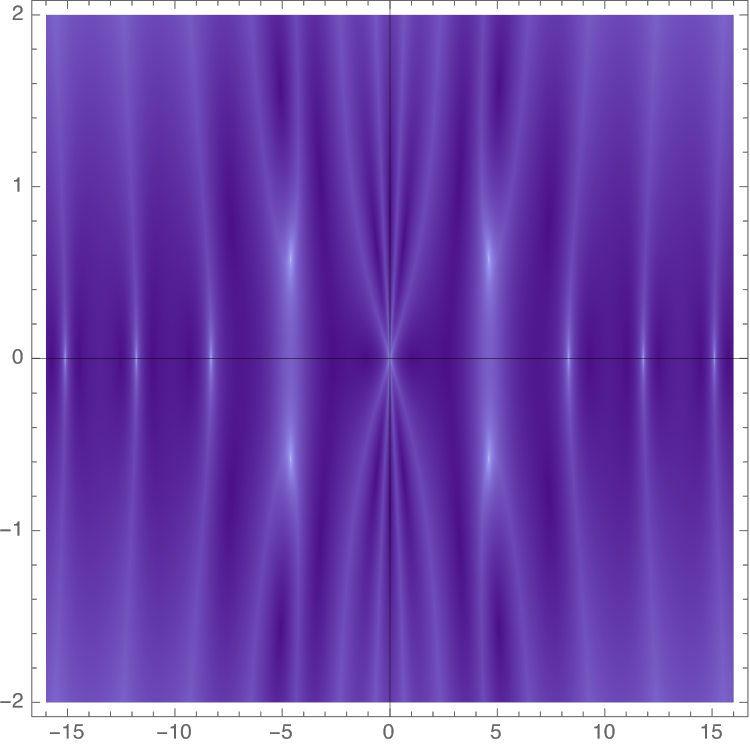} \label{q_4_8i}}                         
                \caption{Density plot of absolute value of $\log \mathcal{F}^{V/S}_l(\mathfrak{r} \, \omega)^2$ in the complex $\mathfrak{r} \, \omega$ plane for $l=2$ and $l=4$. In the scalar case, we further multiply $\mathcal{F}^{S}_l$ by  $|\mathfrak{r} \, \omega|^{-3}$ to highlight the position of the zeros. Note that the scalar function (for both $l$'s) has four zeros that are not real.} \label{fig: omega density plot}
\end{figure}

Next, we impose the conformal boundary conditions. As a consequence of the gauge \eqref{eqn: sphere l>2 gauge fixing}, the conformal boundary conditions \eqref{eqn: bdry cond 1} and \eqref{eqn: bdry cond 2} become
\begin{equation}\label{eqn: sphere vector bdry cond}
	\frac{\Phi^V}{r} + \partial_r \Phi^V |_{r=\mathfrak{r}} \= 0 \, ,
\end{equation}
\begin{equation}\label{eqn: sphere scalar bdry cond}
	\Bigg(\frac{l(l+1) \left(2l(l+1)-3\right)}{r^4}+\frac{4l(l+1)-4}{r^2} \partial_t^2 + 2 \partial_t^4\Bigg) \Phi^S + \Bigg(\frac{3l(l+1)-4}{r^2}+2\partial_t^2\Bigg)\frac{\partial_r\Phi^S}{r} \Big|_{r=\mathfrak{r}} \= 0 ,
\end{equation}
where we have imposed the equation of motion \eqref{eqn: sphere master field} in order to eliminate terms containing $\partial_r^2 \Phi^S$. Since the two master fields $\Phi^V$ and $\Phi^S$ are decoupled both in  the equation of motion \eqref{eqn: sphere master field} as well as the boundary conditions \eqref{eqn: sphere vector bdry cond} and \eqref{eqn: sphere scalar bdry cond}, they can be analysed independently. 
\newline\newline
\textbf{Vector perturbation.} For the vector perturbation, by using in the general solution \eqref{eqn: spherical l>2 sol}, the boundary condition \eqref{eqn: sphere vector bdry cond} gives rise to the following transcendental equation for $\mathfrak{r} \, \omega$:
\begin{equation}\label{eqn: F^V = 0}
	\mathcal{F}^V_l(\mathfrak{r} \, \omega) \,\equiv\, (l+2)J_l(\mathfrak{r} \, \omega) - \mathfrak{r} \, \omega J_{l+1}(\mathfrak{r} \, \omega) \= 0 \, .
\end{equation}
This equation restricts the allowed values of the frequencies $\omega$ of the vector perturbation to a discrete set, that we call $\Omega_{V,l}$. We show those cases for $l=2,4$ in figures \ref{q_2_4} and \ref{q_2_4i}, respectively.
 
Given the set $\Omega_{V,l}$, one can write down the general solution of master field $\Phi^V$ for a fixed $l \geq 2$ as
\begin{equation}\label{eqn: sol vector perturbation}
	\Phi^V(t,r) \= \sum_{\omega \in \Omega_{V,l}} \alpha_l^V (\mathfrak{r} \, \omega) r\omega J_l (r\omega) e^{-i \omega t} + \text{c.c.} 
\end{equation}
Plugging back into \eqref{eqn: spherical l>2 ansatz} and setting $\Phi^S\=0$, we obtain a general gauge-fixed vector perturbation for a given angular momentum $l\geq 2$ and preserving the conformal boundary data of the timelike tube at $r\=\mathfrak{r}$. We observe that the solution does not affect the background Weyl factor on the timelike tube.
\newline\newline
\textbf{Scalar perturbation.} Similarly, inserting the solution \eqref{eqn: spherical l>2 sol} into the boundary condition for the scalar perturbation \eqref{eqn: sphere scalar bdry cond}, we obtain the following condition for $\mathfrak{r} \, \omega$,
\begin{multline} \label{eqn: F^S = 0}
	\mathcal{F}^S_l(\mathfrak{r} \, \omega) \equiv \left(-4-4l+5l^2+7l^3+2l^4-2(l(3+2l)-1)\mathfrak{r}^2 \omega^2 + 2 \mathfrak{r}^4 \omega^4\right) J_l(\mathfrak{r} \, \omega)  \\ + \mathfrak{r} \, \omega \left(4-3l(l+1)+2\mathfrak{r}^2 \omega^2\right) J_{l+1}(\mathfrak{r} \, \omega)   = 0 \, . 
\end{multline}
Let $\Omega_{S,l}$ be the set of zeros of $F^S_l(\mathfrak{r} \, \omega)$, then the general solution for the master field $\Phi^S$ for a fixed $l \geq 2$ is given by
\begin{equation}\label{eqn: sol scalar perturbation}
	\Phi^S(t,r) \= \sum_{\omega \in \Omega_{S,l}} \alpha_l^S (\mathfrak{r} \, \omega) r\omega J_l (r\omega) e^{-i \omega t} + \text{c.c.} 
\end{equation}
Plugging back into \eqref{eqn: spherical l>2 ansatz} and setting $\Phi^V\=0$, we obtain a general gauge-fixed scalar perturbation for a given angular momentum $l\geq 2$ and preserving the conformal boundary data of the timelike tube at $r\=\mathfrak{r}$. A scalar perturbation of fixed $l$ gives rise to a non-vanishing Weyl factor on the timelike tube,
\begin{eqnarray}
	 && h_{\mu \nu}dx^\mu dx^\nu |_{r=\mathfrak{r}} \=  \\
	 &&  \text{Re} \left[ \sum_{\omega \in \Omega_{S,l}} \alpha_l^S (\mathfrak{r} \, \omega) \, \omega \Bigg(2 \, \mathfrak{r} \, \omega J_{l+1}(\mathfrak{r} \, \omega) - \left(2+l(l+3)-2 \, \mathfrak{r}^2\omega^2\right)J_l(\mathfrak{r} \, \omega)\Bigg) \, e^{-i \omega t} \, \mathbb{S}    \right] \left(-dt^2 + \mathfrak{r}^2 d \Omega^2\right) \, . \nonumber
\end{eqnarray}
 
\begin{figure}[h!]
        \centering
         \subfigure[$l = 2$]{
                \includegraphics[scale=0.55]{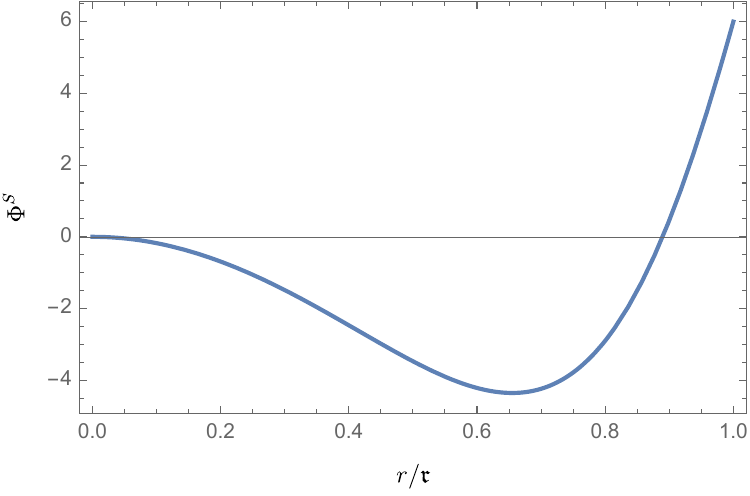}\label{fig_a}}  \quad\quad
                 \subfigure[$l = 4$]{
                \includegraphics[scale=0.55]{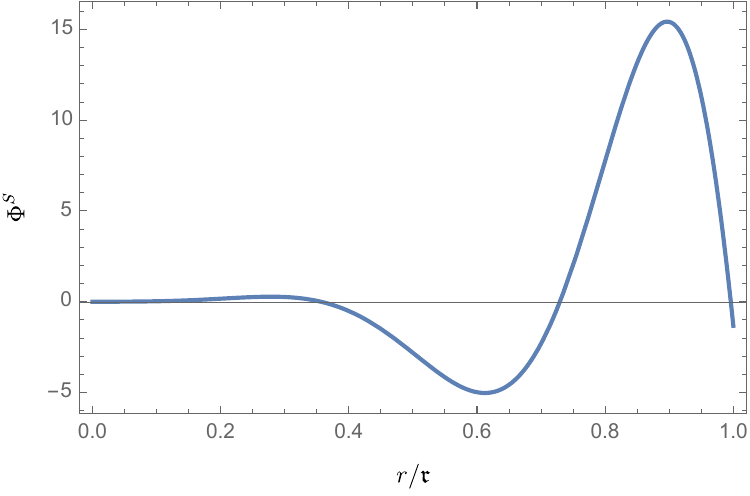} \label{fig_b}}  \quad\quad        
                \caption{Radial profile of $\Phi^S$ as a function of $r/\mathfrak{r}$, at fixed time $t/\mathfrak{r} = 1$ and angular momentum $l$. Only the two frequencies that have positive imaginary part are shown in the plots. Profiles are smooth for any time $t/\mathfrak{r}$.} \label{fig: radial}
\end{figure}

Unlike the set of frequencies for the vector perturbation $\Omega_{V,l}$, the scalar perturbation modes contain two pairs of complex conjugate frequencies. These frequencies can be computed numerically. For the lower angular momenta, for instance, we obtain
\begin{equation}
	\mathfrak{r} \, \omega \=
	\begin{cases}
		\pm 2.50459 \pm 0.413246 i \, ,&  l=2 \, , \\
		\pm 4.60402 \pm 0.579649 i \, ,&  l=4 \, , \\
		\pm 6.64021 \pm 0.664807 i \, ,&  l=6 \, .
	\end{cases}
\end{equation}
See figures \ref{q_4_8} and \ref{q_4_8i} for density plots for $l=2,4$, respectively. In the large-$l$ limit, we numerically find that the four complex frequencies behave roughly as $\mathfrak{r}\omega \approx \pm l \pm i \, l^{1/3}$. For modes with positive imaginary part, the corresponding gravitational perturbation grows exponentially in time. As such,  the perturbative method  breaks down at late times, and a non-linear analysis is required. At any given time though, the radial profile is smooth, see figure \ref{fig: radial}. The situation here is in  contrast to the problem with Dirichlet boundary conditions where it was shown in \cite{Andrade:2015gja} that the interior modes in flat space are linearly stable. On the other hand, it might be interesting to compare the effect to the AdS instability discussed in \cite{Bizon:2011gg}.

We note that asymptotic solutions to \eqref{eqn: F^V = 0} and \eqref{eqn: F^S = 0} at large real values of $\mathfrak{r} \, \omega$ can be obtained analytically. For fixed $l\geq2$, they are given by
\begin{equation}
	\mathfrak{r} \, \omega \= \begin{cases}
	\pm \frac{\pi}{2} \left(2n - l-1 \right)   &  \text{,}  \quad \text{vector perturbation}   \\
	\pm \frac{\pi}{2} \left( 2n-l \right)    & \text{,} \quad \text{scalar perturbation} 
	\end{cases} \,, \qquad
	n \in \mathbb{N} \,.
\end{equation}

\textbf{Physical diffeomorphism.} We now consider a gravitational perturbation which can be written as a pure diffeomorphism, 
\begin{equation}
	h_{\mu \nu} \= \nabla_\mu \xi_\nu + \nabla_\nu \xi_\mu \, , \qquad \qquad \xi_\mu dx^\mu \= \mathcal{T}_t \mathbb{S} dt + \mathcal{T}_r \mathbb{S} dr + \left(\mathcal{L}^S \mathbb{S}_i + \mathcal{L}^V \mathbb{V}_i \right) r d\Omega^i \, ,
\end{equation}
for $l\geq 2$. The conformal boundary conditions \eqref{eqn: diffeo bdry cond 1} and \eqref{eqn: diffeo bdry cond 2} impose that
\begin{eqnarray}
\begin{cases}
	\left. 2r\partial_t \mathcal{T}_t  + \sqrt{l(l+1)}\mathcal{L}^S+2\mathcal{T}_r  \right|_{r=\mathfrak{r}} & \=0  \,, \\
	\left. r\partial_t \mathcal{L}^S - \sqrt{l(l+1)}\mathcal{T}_t \right|_{r=\mathfrak{r}} & \= 0 \,, \\
	\left. \mathcal{L}^S \right|_{r=\mathfrak{r}} & \= 0  \,, \\
	\left. \left(-r^2\partial_t^2 + 2-l(l+1) \right)\mathcal{T}_r  \right|_{r=\mathfrak{r}} & \= 0  \,, \\
	\left. \mathcal{L}^V \right|_{r=\mathfrak{r}} & \= 0  \,,
	\end{cases}
\end{eqnarray}
for $l\geq 2$. It follows that $\mathcal{L}^S|_{r=\mathfrak{r}} \= \mathcal{L}^V|_{r=\mathfrak{r}}\=0$. Then, the remaining conditions become
\begin{equation}
	\left. r \, \partial_t \mathcal{T}_t + \mathcal{T}_r \right|_{r=\mathfrak{r}}\= 0 \, , \qquad \left.\mathcal{T}_t\right|_{r=\mathfrak{r}} \= 0 \,, \qquad \left. \left(-r^2 \partial_t^2 + \left( 2-l(l+1)\right) \right)\mathcal{T}_r  \right|_{r=\mathfrak{r}}  \= 0 \, .
\end{equation}
These conditions give rise to $\mathcal{T}_t |_{r=\mathfrak{r}} \= \mathcal{T}_r |_{r=\mathfrak{r}} \= 0$. Clearly, the last condition is equivalent to the condition that the given diffeomorphism does not move the location of the timelike boundary \eqref{eqn: diffeo bdry cond perp}. Therefore, any linearised diffeomorphism $\xi_\mu$ with $l \geq 2$ and preserving the conformal boundary data of the timelike tube is a trivial diffeomorphism which can be gauged away.
 
We note that since for this argument we have not imposed any gauge condition, the result applies to any gauge choice.
\newline\newline
\textbf{Brief remark on outgoing modes.} Though we are not focusing on the modes exterior to the wordtube that propagate toward null infinity, we would like to make a brief comment. The structure of the linearised solutions is as for the interior modes. The difference is that we no longer need to impose that the modes are non-singular at the origin since that region is excised from the spacetime.\footnote{This leads to the possibility of including a portion of the negative mass Schwarzschild solution, which would have a naked timelike singularity at the origin.} Near null infinity we will generically have a linear combination of incoming and outgoing spherical waves. We further imagine imposing that the waves are purely outgoing for each angular momentum $l$. The essential difference between the exterior and interior modes is that once an exterior mode is outgoing, it will eventually reach null infinity rather than returning back.

As an example, we consider the fixed-$l$ scalar solution
\begin{equation}
\Phi^S(t,r) =  \sum_\omega \alpha_l^S (\mathfrak{r} \, \omega) \, r  \omega  \left( J_l (r\omega) + i \, Y_l(r\omega) \right) e^{-i \omega t} + \text{c.c.}  \,,
\end{equation}
that is purely outgoing in the radial direction. To preserve the conformal boundary conditions, we can restrict to initial data on $\Sigma$, so that the purely outgoing modes have support away from the intersection of $\Sigma$ with $\Gamma$. 

More general configurations will also have ingoing modes and in such a situation one must impose the boundary condition (\ref{eqn: sphere scalar bdry cond}) at $r=\mathfrak{r}$. The condition is a mild generalisation of (\ref{eqn: F^S = 0}), since we are also allowed to have modes built from the $Y_l(r\omega)$. Nevertheless, at least for the modes built exclusively from the $J_l(r\omega)$, the boundary condition is equivalent to (\ref{eqn: F^S = 0}), and consequently, a subset of configurations will also exhibit exponential growth. As another example, let us consider the purely outgoing condition equipped with conformal boundary conditions at $r=\mathfrak{r}$, and take $l=2$. In this case, we find two growing modes with values $\mathfrak{r}\omega= \pm 1.78355 + 0.372158 i$. 

\subsection{Modifying $\delta K$}

We now consider conformal boundary conditions for which the trace $K$ of the extrinsic curvature is varied slightly away from $2/\mathfrak{r}$ while the conformal class of the boundary metric remains intact. This is achieved through the following boundary conditions on the metric perturbation:
\begin{equation}\label{eqn: spherical modified K}
	h_{mn}-\frac{1}{3}\bar{g}_{mn} h^p{}_p |_{r=\mathfrak{r}} \= 0 \, , \qquad \partial_r h^m{}_m - 2 \mathcal{D}^m h_{rm}- \frac{2}{\mathfrak{r}}h_{rr} |_{r=\mathfrak{r}} \= 2\delta K(t,\theta,\phi) \, ,
\end{equation}
where $\delta K (t,\theta,\phi)$ is an arbitrary function of the boundary coordinates so that the trace of the extrinsic curvature of the timelike boundary becomes $K = {2}/{\mathfrak{r}}+ \epsilon \, \delta K (t,\theta,\phi)$.

In order to find metric perturbation obeying modified boundary conditions, we decompose $\delta K$ using the spherical harmonic functions $\mathbb{S}$,
\begin{equation}
	\delta K (t,\theta,\phi) \= \sum_{l=0}^\infty \sum_{m=-l}^l \delta K_{l,m}(t) \mathbb{S}(\theta,\phi) \, .
\end{equation}
We then consider metric perturbations for each $l$ separately. Since the $m$-dependence is not important for the following calculation, we will neglect it for the rest of the section.
\subsubsection{$l=0$ mode}
We first consider the modified boundary conditions in the spherically symmetric sector. Specifically, we consider the following gravitational perturbation:
\begin{equation}
	h_{\mu \nu} \= \nabla_\mu \xi_\nu + \nabla_\nu \xi_\mu \, ,  \qquad \qquad \xi_\mu dx^\mu \= \xi_t(t,r) dt + \xi_r(t,r) dr \, ,
\end{equation}
where $\xi_t(t,r)$ and $\xi_r (t,r)$ are some arbitrary functions of $t$ and $r$. This perturbation automatically satisfies the linearised Einstein field equation as it is a pure diffeomorphism. Working in harmonic gauge imposes that it satisfies \eqref{eqn: l=0 eom of diffeo}.

Imposing the boundary conditions \eqref{eqn: spherical modified K}, we find
\begin{equation}\label{eqn: spherical modified K bdry l=0}
	\partial_t \xi_t + \frac{\xi_r}{r} |_{r=\mathfrak{r}} \= 0 \, , \qquad \left(\partial_t^2 - \frac{2}{r^2}\right) \xi_r |_{r=\mathfrak{r}}\= \delta K_{l=0}(t) \, .
\end{equation}
Viewing the second boundary condition as a differential equation on the boundary value $ \xi_r |_{r=\mathfrak{r}}$, one can write down the general solution as
\begin{equation}\label{eqn: modified spherical bdry xi_r}
	\xi_r |_{r=\mathfrak{r}} \= \left(\alpha^{(+)} + \mathfrak{r} \int_0^t dt' e^{-\sqrt{2} t'/\mathfrak{r}} \frac{\delta K_{l=0}(t')}{2\sqrt{2}}\right)e^{\sqrt{2}t/\mathfrak{r}} + \left(\alpha^{(-)} - \mathfrak{r}\int_0^t dt' e^{\sqrt{2} t'/\mathfrak{r}} \frac{\delta K_{l=0}(t')}{2\sqrt{2}}\right)e^{-\sqrt{2}t/\mathfrak{r}} \, , 
\end{equation}
where $\alpha^{(+)}$ and $\alpha^{(-)}$ are reminiscent of the exponentially growing and decaying modes found in section \ref{sec: l=0 time dependent}. We then use the first boundary condition from \eqref{eqn: spherical modified K bdry l=0} to obtain the boundary value of the time component $ \xi_t |_{r=\mathfrak{r}}$. Lastly, we use the harmonic gauge constraints \eqref{eqn: l=0 eom of diffeo} to radially extend the solution into the interior and calculate the metric perturbation as $h_{\mu\nu}= \nabla_\mu \xi_\nu + \nabla_\nu \xi_\mu$.

From \eqref{eqn: modified spherical bdry xi_r}, it can be seen that turning on $\delta K_{l=0} (t)$ for some finite time interval $\left[ t_i , t_f \right]$, can result in changing the exponentially growing/decaying behaviour of the solution. Thus, there is a sense in which the exponentially growing mode $\alpha^{(+)}$ can be tamed by turning on an appropriate $\delta K_{l=0}(t)$ function.

\textbf{Example.} Consider a Dirac delta function located at some time $t_0>0$ with strength $\kappa$, $\delta K_{l=0}(t) \= \kappa \, \delta (t-t_0)$. Expression \eqref{eqn: modified spherical bdry xi_r} now becomes
\begin{equation}
	\xi_r |_{r=\mathfrak{r}} \= \left(\alpha^{(+)} + \kappa \, \mathfrak{r} \frac{e^{-\sqrt{2} t_0/\mathfrak{r}}}{2\sqrt{2}}\Theta(t-t_0)\right)e^{\sqrt{2}t/\mathfrak{r}} + \left(\alpha^{(-)} - \kappa \, \mathfrak{r} \frac{e^{\sqrt{2} t_0/\mathfrak{r}}}{2\sqrt{2}}\Theta(t-t_0)\right)e^{-\sqrt{2}t/\mathfrak{r}} \, ,
\end{equation}
where $\Theta(t)$ is the Heaviside step function. By tuning $\kappa \, \mathfrak{r} \= -2\sqrt{2} e^{\sqrt{2}t_0} \alpha^{(+)}$, we find that
\begin{equation}\label{eqn: modified spherical bdry xi_r 2}
	\xi_r |_{r=\mathfrak{r}} \= \alpha^{(+)}\left(1-\Theta(t-t_0)\right)e^{\sqrt{2} t/\mathfrak{r}} + \left(\alpha^{(-)} + \alpha^{(+)} e^{2\sqrt{2}t_0/\mathfrak{r}}\Theta(t-t_0)\right) e^{-\sqrt{2}t/\mathfrak{r}} \, .
\end{equation}
This solution does not exhibit exponential growth at late times anymore. In fact, it is exponentially decaying after time $t_0$. Given that there is no exponentially growing mode for $l=1$, we now directly analyse the case $l\geq 2$.
\subsubsection{$l\geq2$ modes} Recall that upon fixing the gauge \eqref{eqn: sphere l>2 gauge fixing}, the general metric perturbation for fixed $l$ can be written in terms of the master fields $\Phi^V$ and $\Phi^S$, see \eqref{eqn: spherical l>2 ansatz}. The master fields satisfy \eqref{eqn: sphere master field}, and hence the general solution is given by \eqref{eqn: spherical l>2 sol}. Regularity of the solution in the interior still requires $\beta^{V/S} = 0$.

Imposing the modified boundary conditions \eqref{eqn: spherical modified K} leads to
\begin{equation}
	\frac{\Phi^V}{r} + \partial_r \Phi^V |_{r=\mathfrak{r}} \= 0 \, ,
\end{equation}
\begin{equation}\label{eqn: spherical modified K l >= 2}
	\Bigg(\frac{l(l+1) \left(2l(l+1)-3\right)}{r^4}+\frac{4l(l+1)-4}{r^2} \partial_t^2 + 2 \partial_t^4\Bigg) \Phi^S + \Bigg(\frac{3l(l+1)-4}{r^2}+2\partial_t^2\Bigg)\frac{\partial_r\Phi^S}{r} \Big|_{r=\mathfrak{r}} \= -\frac{4\delta K_{l}(t)}{\mathfrak{r}^2}\, .
\end{equation}
Since the vector master field $\Phi^V$ obeys the same boundary condition, its solution is the same as in the previous subsection. 

To solve the boundary condition \eqref{eqn: spherical modified K l >= 2}, we apply the time-Fourier transform 
\begin{equation}
\delta K_{l}(t) \= \int_\mathbb{R} \frac{d \omega}{2 \pi} e^{-i \omega t}\delta K_{l}(\omega) \, ,
\end{equation}
after which the boundary condition \eqref{eqn: spherical modified K l >= 2} becomes
\begin{equation}
	\mathcal{F}^S_l(\mathfrak{r} \, \omega)\alpha^S (\mathfrak{r}\, \omega) \= - \frac{4 \delta K_l(\omega)}{\mathfrak{r}^2} \, .
\end{equation}
This  fixes $\alpha^S(\mathfrak{r} \, \omega)$ in terms of $\delta K_l(\omega)$. Inserting this back into \eqref{eqn: spherical l>2 sol} and integrating over $\omega$, one finds that the inhomogeneous part of the solution is given by
\begin{equation}
	\Phi^S(t,r) \= - \int_\mathcal{C} \frac{d \omega}{2\pi} \frac{4 \, \delta K_l (\omega)}{\mathfrak{r}^2 \mathcal{F}^S_l (\mathfrak{r} \, \omega)}  r \omega J_l(r \omega) e^{-i \omega t} + \text{c.c.}\, , 
\end{equation}
where $\mathcal{C}$ is a contour in the complex $\omega$-plane chosen such that it lies above all the zeros of $\mathcal{F}^S_l(\mathfrak{r} \, \omega)$. To write down a complete solution, one must include a homogeneous solution which is $\Phi^S$ obeying the boundary condition with $\delta K_l \= 0$. So, the general $\Phi^S$ preserving the conformal class of the spherical boundary and having $K \= \frac{2}{\mathfrak{r}}+\delta K_l(t)\,\mathbb{S}(\theta,\phi)$ for fixed $l\geq 2$ is given by
\begin{equation}\label{eqn: Phi^S integrated over contour}
	\Phi^S(t,r) \= - \int_\mathcal{C} \frac{d \omega}{2\pi} \frac{4 \, \delta K_l (\omega)}{\mathfrak{r}^2 \mathcal{F}^S_l (\mathfrak{r} \, \omega)}  r \omega J_l(r \omega) e^{-i \omega t} + \sum_{\omega' \in \Omega_{S,l}} \alpha_l^S (\mathfrak{r} \, \omega') r\omega' J_l (r\omega') e^{-i \omega' t} + \text{c.c.}\, , 
\end{equation}
where $\alpha^S_l (\mathfrak{r}\, \omega')$ is an arbitrary function.

\textbf{Example.} Consider $\delta K_{l}(t) \= \kappa_l \, \delta(t-t_0)$, with $\kappa_l$ constant. Its Fourier transform is given by $\delta K_l (\omega) \= \kappa_l e^{i \omega t_0}$. Plugging this into \eqref{eqn: Phi^S integrated over contour}, we find
\begin{equation}
	\Phi^S (t,r) \= - \int_\mathcal{C} \frac{d \omega}{2\pi} \frac{4 \, \kappa_l \, e^{-i \omega(t-t_0)}}{\mathfrak{r}^2\mathcal{F}_l^S(\mathfrak{r} \, \omega)} r \omega J_{l} (r\omega) + \sum_{\omega' \in \Omega_{S,l}} \alpha_l^S (\mathfrak{r} \, \omega') r\omega' J_l (r\omega') e^{-i \omega' t} + \text{c.c.}\, , 
\end{equation}
When $t<t_0$, the contour can be closed in the upper half plane. Since there is no pole lying above the contour, the integral simply leads to zero. As a result, information from turning on $\delta K_l(t)$ at time $t_0$ does not propagate back in time. For time $t>t_0$, the contour can be closed in the lower half plane. Using the residue theorem, one finds that the integral reduces to a sum over the zeros of $\mathcal{F}^S_l(\mathfrak{r}\, \omega)$. Combining with the homogeneous part of the solution, one arrives at
\begin{equation}
	\Phi^S (t,r) \= \sum_{\omega \in \Omega_{S,l}} \left(\frac{4 i \kappa_l e^{i \omega t_0}}{\mathfrak{r}^2 \left(\mathcal{F}^S_l (\mathfrak{r} \, \omega) \right)' } \Theta(t-t_0) + \alpha_l^S(\mathfrak{r} \, \omega) \right)r \omega J_l (r\omega)e^{-i\omega t} + \text{c.c.} \, ,
\end{equation}
where $\left( \mathcal{F}^S_l (x) \right)' \,\equiv\, \left.\frac{d\mathcal{F}^S_l}{dx}\right|_x$. The resulting solution implies that turning on $\delta K_l(t)$ shifts the amplitude of the gravitational modes $\alpha^S_l(\mathfrak{r} \, \omega)$ in the bulk by a factor that depends on $\kappa_l$. It follows that a gravitational mode associated to a frequency with positive imaginary part can be eliminated by choosing appropriate parameters $\kappa_l$ and $t_0$. If there are more than one of these frequencies, then a slightly more involved $\delta K_l (t)$ would be needed. 

As a consequence, and similarly to the spherically symmetric sector case, we can judiciously select a $\delta K(t,\theta,\phi)$ for some finite time interval such that the metric perturbation does not exhibit exponential growth at late times.

\subsection{Dirichlet spherical boundary}

In this section, we consider Dirichlet boundary conditions instead and show that, in this case, the geometric uniqueness of the linearised Einstein equation is preserved.

Linearised gravitational modes  subject to Dirichlet boundary conditions which are not local diffeomorphisms were studied in \cite{Andrade:2015gja}, using the Kodama-Ishibashi formalism. 
Here, we study metric perturbations that are locally diffeomorphisms. These take the form $h_{\mu \nu} \= \nabla_\mu \xi_\nu + \nabla_\nu \xi_\mu $, where
\begin{equation}\label{eqn: spherical Dirichlet ansatz}
	\xi_\mu dx^\mu \= 
	\begin{cases}
	\mathcal{T}_t \, \mathbb{S} \, dt + \mathcal{T}_r \, \mathbb{S} \, dr  &, \qquad l\=0 \, ,  \\
	\mathcal{T}_t \, \mathbb{S} \, dt + \mathcal{T}_r \, \mathbb{S} \, dr + \left(\mathcal{L}^S \, \mathbb{S}_i + \mathcal{L}^V \mathbb{V}_i \right) r d\Omega^i \, &, \qquad l \geq 1 \, .
	\end{cases}
\end{equation}
Working in harmonic gauge requires that
\begin{equation}
	\nabla^\nu \nabla_\nu \xi_\mu \= 0\, . 
\end{equation}
 
The induced metric on the timelike boundary is given by 
\begin{equation}
	\left. ds^2 \right|_{r=\mathfrak{r}} \= -dt^2  +\mathfrak{r}^2 d\Omega^2 \, . 
	\end{equation} 
 
Instead of imposing conformal boundary conditions on this timelike boundary, we impose Dirichlet boundary conditions,
\begin{equation}\label{eqn: spherical Dirichlet bdry cond}
	\left. h_{m n} \right|_{r=\mathfrak{r}} \= 0 \, .
\end{equation}
Inserting the ansatz \eqref{eqn: spherical Dirichlet ansatz} into the boundary conditions \eqref{eqn: spherical Dirichlet bdry cond}, one obtains a set of boundary conditions in terms of $\mathcal{T}_t$, $\mathcal{T}_r$, $\mathcal{L}^S$, and $\mathcal{L}^V$, that depend on $l$. We are interested in physical diffeomorphisms, so we require that $\xi_\mu$ does not obey the condition \eqref{eqn: diffeo bdry cond perp}, and hence
\begin{equation}\label{eqn: Dirichlet sphere physical diffeo cond}
	\left. \mathcal{T}^r \right|_{r=\mathfrak{r}} \neq 0 \, .
\end{equation}
 
Using the metric perturbation constructed from \eqref{eqn: spherical Dirichlet ansatz} that satisfies both \eqref{eqn: spherical Dirichlet bdry cond} and \eqref{eqn: Dirichlet sphere physical diffeo cond}, we now show that it is not possible to build waves from such solutions which have compact support outside the timelike tube at some initial time while subsequently moving into the interior of the tube.
 
Let us first consider the $l=0$ case, the boundary conditions \eqref{eqn: spherical Dirichlet bdry cond} become 
\begin{eqnarray}
\begin{cases}
	\left. \partial_t \mathcal{T}_t \right|_{r=\mathfrak{r}} & \=0  \,, \\
	\left. \mathcal{T}_r \right|_{r=\mathfrak{r}} & \= 0 \, .
	\end{cases}
\end{eqnarray}
It is straightforward to see that the second condition contradicts \eqref{eqn: Dirichlet sphere physical diffeo cond}. Hence, there is no physical diffeomorphism with $l=0$.
 
For $l=1$, the boundary conditions \eqref{eqn: spherical Dirichlet bdry cond} become
\begin{eqnarray}
\begin{cases}
	\left. \partial_t \mathcal{T}_t \right|_{r=\mathfrak{r}} & \=0  \,, \\
	\left. r\partial_t \mathcal{L}^S - \sqrt{2}\mathcal{T}_t \right|_{r=\mathfrak{r}} & \= 0 \,, \\
	\left. \mathcal{L}^S +\sqrt{2}\mathcal{T}_r  \right|_{r=\mathfrak{r}} & \= 0  \,.
	\end{cases}
\end{eqnarray}
Imposing the condition \eqref{eqn: Dirichlet sphere physical diffeo cond}, one finds that the solution must either be constant\footnote{This solution was studied in \cite{Andrade:2015gja}, where it was shown using symplectic structure in the phase space that it is unphysical when no black hole is present.} or linear in time. By radially extending this to the bulk, a physical diffeomorphism with $l\=1$ which preserves the Dirichlet boundary data is obtained. However, as the resulting diffeomorphism linearly grows in time, it is fixed once we impose the initial condition.
 
Lastly, we consider solutions with $l\geq2$. The Dirichlet boundary conditions \eqref{eqn: spherical Dirichlet bdry cond} now impose that 
\begin{eqnarray}
\begin{cases}
	\left. \partial_t \mathcal{T}_t \right|_{r=\mathfrak{r}} & \=0  \,, \\
	\left. r\partial_t \mathcal{L}^S - \sqrt{l(l+1)}\mathcal{T}_t \right|_{r=\mathfrak{r}} & \= 0 \,, \\
	\left. \mathcal{L}^S \right|_{r=\mathfrak{r}} & \= 0  \,, \\
	\left. \sqrt{l(l+1)}\mathcal{L}^S + 2\mathcal{T}_r  \right|_{r=\mathfrak{r}} & \= 0  \,, \\
	\left. \mathcal{L}^V \right|_{r=\mathfrak{r}} & \= 0  \,.
	\end{cases}
\end{eqnarray}
Note that combining the third and fourth boundary conditions above, contradicts  \eqref{eqn: Dirichlet sphere physical diffeo cond}. Hence, there is no physical diffeomorphism with $l\geq2$ \cite{Andrade:2015fna}.
 
In summary, when $l=0$ and $l\geq2$, there is no metric built from a pure diffeomorphism satisfying both Dirichlet boundary conditions \eqref{eqn: spherical Dirichlet bdry cond} and the condition for being physical \eqref{eqn: Dirichlet sphere physical diffeo cond}, at the same time. For $l=1$, there exists such solutions, however the boundary conditions force them to grow linearly in time which implies that they are uniquely fixed once the initial conditions are specified. 

As a result, and in contrast to the standard corner analysis of \cite{An:2021fcq,Anninos:2022ujl}, geometric uniqueness appears to be preserved for spherical boundaries with Dirichlet boundary conditions. It may be interesting to relate this observation to the improved Euclidean Dirichlet problem in those cases where the extrinsic curvature obeys certain positivity conditions \cite{Anderson:2006lqb,Witten:2018lgb}. In any case, it is tempting to conjecture that for the case of a spherical  spatial boundary, the gravitational Dirichlet problem is non-linearly well-posed, but further analysis is required.

\section{Black hole conformal thermodynamics}\label{sec: euclidean bh}

The problem of a Euclidean black hole in a finite size box with fixed induced metric at the boundary $\Gamma$ was considered in \cite{York:1986it} by York. In this section,  we consider an extension of the York setup for a finite size space subject to the conformal boundary conditions of \cite{An:2021fcq}. Aspects of this problem were also considered in \cite{Odak:2021axr}. The purpose of the section is to assess how the thermodynamic quantities of the black hole are affected by the alternative choice of boundary conditions.

Concretely, we consider a Euclidean black hole solution preserving the following boundary data,
\begin{equation}
	ds^2 |_\Gamma \= e^{2 \boldsymbol{\omega}} \left(d\tau^2 + \mathfrak{r}^2 d\Omega^2\right) \, , \qquad \qquad K \= \text{constant} \, ,
\end{equation}
for some unspecified function $\boldsymbol{\omega}$. We take the topology of $\Gamma$ to be $S^1 \times S^2$.  The Euclidean time coordinate $\tau \sim \tau +\beta$ parameterises the $S^1$ factor. The parameter $\mathfrak{r}$ characterises the size of the $S^2$. Given that only the conformal structure of the boundary $\Gamma$ is specified, only the dimensionless parameter $\beta / \mathfrak{r}$ is geometrically meaningful. The trace of the extrinsic curvature $K$ has dimensions of inverse length, which provides a length scale for the problem.
 
\subsection{Black hole solution} 
 
Assuming stationarity, the Euclidean black hole solution obeying the given boundary data is 
\begin{equation}\label{bhs}
ds^2 \= \frac{e^{2 \boldsymbol{\omega}}}{1-\frac{2M}{r}}dr^2 + e^{2 \boldsymbol{\omega}}\left(\frac{1-\frac{2M}{r}}{1-\frac{2M}{\mathfrak{r}}}d\tau^2 + r^2 d\Omega^2\right) \, ,
\end{equation}
where the Weyl factor is a constant given by
\begin{equation}
	e^{\boldsymbol{\omega}} \= \frac{1-\frac{3M}{2\mathfrak{r}}}{\sqrt{1-\frac{2M}{\mathfrak{r}}}}\frac{2}{K\mathfrak{r}} \, .
\end{equation}
The boundary $\Gamma$ is located at $r=\mathfrak{r}$. To map the solution to more standard Schwarzschild coordinates, one can perform the following coordinate transformation:
\begin{equation}
	r \rightarrow r_\text{Sch} \= e^{\boldsymbol{\omega}} r \, \quad , \quad \,
	\tau \rightarrow \tau_\text{Sch} \=  \frac{e^{\boldsymbol{\omega}}\tau}{\sqrt{1-\frac{2M}{\mathfrak{r}}}} \, ,
\end{equation}
where $r_\text{Sch}$ and $\tau_\text{Sch}$ are the Euclidean Schwarzschild coordinates. One thus identify the mass of the black hole and the physical radius of the tube as 
\begin{eqnarray}
	M_\text{bh}  \equiv e^{\boldsymbol{\omega}} M \, \quad , \quad \, 
	\mathfrak{r}_\text{tube} \equiv& e^{\boldsymbol{\omega}} \mathfrak{r} \, .
\end{eqnarray}
The above solution has a conical singularity at the Euclidean horizon $r_\text{Sch} = 2M_\text{bh}$ or $r = 2M$ unless one identifies
\begin{equation}\label{eqn: roots of bh mass}
	\frac{\beta}{\mathfrak{r}} \= \frac{ 8 \pi M}{\mathfrak{r}}\sqrt{1-\frac{2M}{\mathfrak{r}}} \=   \frac{8 \pi M_\text{bh}}{\mathfrak{r}_\text{tube}}\sqrt{1-\frac{2M_\text{bh}}{\mathfrak{r}_\text{tube}}} \, .
\end{equation}
Since the Weyl factor $e^{\boldsymbol{\omega}}$ is a constant, the parameter $\beta/\mathfrak{r}$ can be interpreted as a ratio between the physical radii of $S^1$ and $S^2$ of the boundary. This equation implies that the dimensionless parameter $\beta/\mathfrak{r}$ controls the size of the black hole relative to the tube. Following \cite{York:1986it,Brown:1992br}, there are two different black hole solutions for a given $\beta/\mathfrak{r}$ if 
\begin{equation}\label{eqn: upper bound of beta/R}
	\frac{\beta}{\mathfrak{r}} \leq \frac{8 \pi}{\sqrt{27}} \, ,
\end{equation}
and only one black hole solution when the equality holds. When $\beta/\mathfrak{r}$ violates the inequality, no black hole of real mass exists. We remark that the upper bound of $\beta/\mathfrak{r}$ in which the black hole exists is independent of the value of $K$. We will denote $M_+$ and $M_-$ the larger and smaller positive roots of  \eqref{eqn: roots of bh mass}, respectively.

\subsection{Thermodynamic quantities}

According to the Gibbons-Hawking prescription for black hole thermodynamics, one can evaluate the on-shell action on the Euclidean black hole solution to compute the leading contribution to the thermal partition function in a canonical ensemble \cite{Gibbons:1976ue}. Here, we would like to consider the problem subject to conformal rather than Dirichlet boundary conditions. To leading order, the modified partition function is related to the Euclidean version of the on-shell action (\ref{eqn: action}) with conformal boundary term  (\ref{IB}) as
\begin{equation}
	\mathcal{Z}\left(\frac{\beta}{\mathfrak{r}},K\right) \approx e^{-I(g^*_{\mu \nu})}~.
\end{equation}

The metric $g_{\mu \nu}^*$ is a classical solution to the Einstein field equation subject to the boundary data $\beta/\mathfrak{r}$ and $K$ on the boundary $\Gamma$ of topology $S^1 \times S^2$. If there are many classical solutions obeying the same boundary data, then we sum all of them when computing $Z$. Since the inverse temperature $\beta$ alone is not fixed but its ratio with the size of the two-sphere $\beta/\mathfrak{r}$ is, we will refer to this as the conformal canonical ensemble. For notational simplicity in what follows we define $\tilde{\beta} \equiv \beta/\mathfrak{r}$.
 
From the previous sub-section, certain regimes in the space  of boundary data permit several classical solutions. Specifically, upon setting $G_N=1$, we find
\begin{eqnarray}
\begin{cases}
	I^{\text{(flat)}} &= -\frac{4 \tilde{\beta}}{3 K^2} \, , \\
	I^{\text{(large bh)}} &= -\frac{4\pi}{3K^2} \frac{M_+}{\mathfrak{r}}\frac{\left(2-\frac{3M_+}{\mathfrak{r}}\right)^3}{1-\frac{2M_+}{\mathfrak{r}}} \, , \\
	I^{\text{(small bh)}} &= -\frac{4\pi}{3K^2} \frac{M_-}{\mathfrak{r}}\frac{\left(2-\frac{3M_-}{\mathfrak{r}}\right)^3}{1-\frac{2M_-}{\mathfrak{r}}} \, . 
	\end{cases}
\end{eqnarray}
We emphasise that $M_\pm/\mathfrak{r}$ should be thought of as a function of $\tilde{\beta}$. The first solution corresponds to the flat Euclidean solution which exists for all values of $\tilde{\beta}$. The latter two correspond to the large and small Euclidean black holes (\ref{bhs}). Both exist only when the inequality \eqref{eqn: upper bound of beta/R} holds, which gives an upper bound for $\tilde{\beta}$. All three Eulidean solutions are defined for any positive value of $K$.
 
Accordingly, one can define the conformal free energy 
\begin{equation}
	\mathcal{F}_\text{conf} \, \equiv \, {I}/{\tilde{\beta}} \, .
\end{equation} 
The Euclidean flat space then has the conformal free energy $\mathcal{F}_\text{conf}^\text{(flat)}=-4/3K^2$ independent of $\tilde{\beta}$. In the low temperature regime, the flat Euclidean solution has the lowest conformal free energy and hence dominates the partition function. At the critical $\tilde{\beta}_c = 32 \pi/27$, the large black hole contribution competes with the flat solution. Then, the large black hole becomes the dominant contribution for high temperatures, $0<\tilde{\beta}<\tilde{\beta}_c$. The small black hole solution is sub-dominant for all $\tilde{\beta}$.
 
One can also define the conformal energy $E_\text{conf}$ and conformal entropy $\mathcal{S}_\text{conf}$ as
\begin{eqnarray}
	E_\text{conf} \, \equiv \, \left. {\partial_{\tilde{\beta}} I} \right|_K \, \quad , \quad \, 
	\mathcal{S}_\text{conf} \, \equiv \, \left. \left(\tilde{\beta} \partial_{\tilde{\beta}} - 1 \right) I \right|_K \, .
\end{eqnarray}
Given these definitions, one can show that all three thermal solutions satisfy a first law relation
\begin{equation}
	\tilde{\beta} \, dE_\text{conf} \=  d \mathcal{S}_\text{conf} \, , \qquad \qquad K \text{ fixed} \, .
\end{equation}
The flat Euclidean solution gives vanishing conformal entropy due to its linear dependence on $\tilde{\beta}$. For both of the black holes, the conformal energy and conformal entropy are given by
\begin{eqnarray}
\begin{cases}
	E_\text{conf} &\= - \frac{4 \left(1-\frac{3M}{2\mathfrak{r}}\right)^2\left(1-\frac{3M}{\mathfrak{r}}\right)}{3K^2 \left(1-\frac{2M}{\mathfrak{r}}\right)^{3/2}}  \= \frac{\left(\frac{3M_\text{bh}}{\mathfrak{r}_\text{tube}}-1\right)\mathfrak{r}_\text{tube}^2}{3\sqrt{1-\frac{2M_\text{bh}}{\mathfrak{r}_\text{tube}}}}  \, , \\
	\mathcal{S}_\text{conf} &\= \frac{16 \pi \left(\frac{M}{\mathfrak{r}}\right)^2 \left(1-\frac{3M}{2\mathfrak{r}}\right)^2}{K^2 \left(1-\frac{2M}{\mathfrak{r}}\right)} \= 4 \pi M_\text{bh}^2 \, .
	\end{cases}
\end{eqnarray}
The final expression of $\mathcal{S}_\text{conf}$ agrees precisely with the Bekenstein-Hawking entropy $A_\text{horizon}/4 G_N$. We note that $\sqrt{1-2M/\mathfrak{r}} \, E_\text{conf}/\mathfrak{r}_\text{tube}$ is equal to the conserved charge in the canonical formalism investigated in \cite{Odak:2021axr}, prior to the addition of a regularisation term.
 
In the high temperature limit or $\tilde{\beta} \rightarrow 0$, the tube approaches the black hole horizon $M/\mathfrak{r} \rightarrow 1/2$, and the entropy scales as
\begin{equation}
	\mathcal{S}_\text{conf} \=  \pi  \left(\frac{3 E_\text{conf}}{K}\right)^{2/3}  \qquad \text{with}  \qquad E_\text{conf} \rightarrow \infty \, .
\end{equation}
 
Finally, one can study the sensitivity of the system as we change $\tilde{\beta}$ while keeping $K$ fixed. This is the heat capacity at constant $K$, which we define as
\begin{equation}
	C_K \, \equiv \, \left. -\tilde{\beta}^2 {\partial_{\tilde{\beta}}^2}  \, I  \right|_K  \, .
\end{equation}
When the large black hole dominates, the heat capacity is given by 
\begin{eqnarray}
	C_K \= - \frac{16 \pi \left(\frac{M_+}{\mathfrak{r}}\right)^2 \left(1-\frac{3M_+}{2\mathfrak{r}}\right)\left(2-\frac{8M_+}{\mathfrak{r}}+\frac{9M_+^2}{\mathfrak{r}^2}\right)}{K^2 \left(1-\frac{2M_+}{\mathfrak{r}}\right)\left(1-\frac{3M_+}{\mathfrak{r}}\right)} \, > \, 0 \, \, \, \, \, \text{for } \, \, \, 0< \tilde{\beta}<\tilde{\beta}_c \, .
\end{eqnarray}
Therefore, when the large black hole dominates the system becomes thermally stable. We also note that in the high temperature limit, the heat capacity diverges quadratically in the temperature, namely
\begin{equation}\label{highT}
	\lim_{ \tilde{\beta} \rightarrow 0} C_K \=  {\frac{8 \pi^3}{G_N K^2 \tilde{\beta}^2}}\,  .
\end{equation}

The thermal phase structure with an unstable small black hole, and a large black hole that dominates over the thermal vacuum at high enough temperatures is reminiscent of the situation for  black holes in anti-de Sitter space \cite{Hawking:1982dh}. Moreover, the quadratic temperature dependence in (\ref{highT}) is similar to that of a large black hole in AdS$_4$ -- or equivalently  a three-dimensional conformal field theory. From this we deduce that the number of putative degrees of freedom scales as $N_{\text{d.o.f.}} \approx \tfrac{1}{G_N K^2}$.\footnote{Note that near the boundary of AdS$_4$ we have $K = 3/\ell$ giving the familiar AdS$_4$/CFT$_3$ expression for $N_{\text{d.o.f.}}$.}  It is straightforward to generalise the high-temperature behaviour to $D$-dimensions, for which the high temperature behaviour goes as $\tilde{\beta}^{2-D}$. The temperature dependence is  different from the high temperature behaviour of a thermally stable black hole in a Dirichlet box \cite{York:1986it} (see also \cite{Alessio:2020lpk,Asante:2021blx}) which instead goes as ${\beta}^2$ for all $D\geq 4$.
\begin{center}
\pgfornament[height=5pt, color=black]{83}
\end{center}
\vspace{5pt}

It is of interest to explore more refined observables subject to conformal boundary conditions in either Lorentzian or Euclidean signature. Of particular interest are  gravitational correlation functions with points anchored on $\Gamma$. The structure of such observables around black hole backgrounds or the vacuum, with vanishing or non-vanishing $\Lambda$, for Lorentzian or Euclidean signature, with the possibility of incorporating matter fields may provide a more quasi-local framework to assess properties of spacetime.

This appears to be a particularly valuable framework for the dynamical characterisation of spacetimes which do not possess an asymptotic region where observables can be naturally anchored. 

For the case of $\Lambda>0$, a more complete characterisation of the dynamical features of the de Sitter horizon  is of particular interest \cite{Anninos:2011zn,Anninos:2018svg,Anninos:2022ujl,Susskind:2021esx,Chapman:2022mqd,Aalsma:2022eru}. According to a theorem of Gao and Wald  \cite{Gao:2000ga}, the Penrose diagram of de Sitter space generally stretches vertically in response to the presence of null-energy preserving excitations. This is in sharp contrast to the behaviour of a black hole, whose Penrose diagram stretches horizontally. Such dynamical effects are tied to the chaotic nature of horizons \cite{Shenker:2013pqa}, now perceived as holographic liquids. Examining how dynamical features of the de Sitter horizon (such as those appearing in the theorem by Gao and Wald) are encoded from a worldtube perspective seems like a  well-motivated exercise.

For the case of $\Lambda<0$, our considerations naturally connect to the AdS/CFT correspondence in the presence of a finite cutoff \cite{McGough:2016lol, Taylor:2018xcy, Hartman:2018tkw, Coleman:2020jte}. 

As a final remark, our worldtube $\Gamma$ might be viewed as an auxiliary part of a more complete picture. Take for instance the Euclidean path-integral for general relativity with $\Lambda>0$,
\begin{equation}
\mathcal{Z}\left[ g_{m n}, K \right] = \int \left[ \mathcal{D} g_{\mu\nu}  \right] \, e^{-I[g_{\mu\nu}]}~,
\end{equation}
on a manifold $\mathcal{M}$ with $S^2\times S^1$ boundary subject to the well-posed conformal data $\{ [g_{m n}]_{\text{conf}},K\}$. We may then ask whether there exists some operation on $\mathcal{Z}$ that retrieves the ordinary Gibbons-Hawking Euclidean $S^4$ path integral \cite{Gibbons:1976ue,Anninos:2020hfj} for general relativity with $\Lambda>0$.  Such an operation may involve path-integrating over  the conformal boundary metric \cite{Alishahiha:2004md,Anninos:2021ydw,Blacker:2023oan}. We have our work cut out for us.

\section*{Acknowledgements}
It is a pleasure to acknowledge Vijay Balasubramanian, Bianca Dittrich, Pau Figueras, Chris Herzog, Diego Hofman, Luis Lehner, Hong Liu, and Beatrix M\"uhlmann for useful discussions. D.A. is funded by the Royal Society under the grant ``The Atoms of a deSitter Universe". The work of D.A.G. is funded by UKRI Stephen Hawking Fellowship ``Quantum Emergence of an Expanding Universe". D.A. and D.A.G. are further funded by STFC Consolidated grant ST/X000753/1. C.M. is funded by STFC under grant number ST/X508470/1.

\appendix

\section{Useful formulae for Gaussian normal coordinates}
\label{app:normal}

In this appendix, we provide some results on geometric quantities using Gaussian normal coordinates
\begin{equation}
	ds^2 \= dx_\perp^2 + \bar{g}_{mn} dx^m dx^n \, ,
\end{equation}
 that  are useful to derive some of the expressions in section \ref{sec:framework}. For instance, the extrinsic curvature in Gaussian normal coordinates is just given by,
\begin{equation}
	K_{m n} \= \frac{1}{2}\partial_\perp \bar{g}_{mn} \, .
\end{equation}
Christoffel symbols with indices in the perpendicular direction are simply given by, 
\begin{equation}
	\Gamma^\perp_{mn} \= - K_{mn} \, , \qquad \qquad \Gamma^m_{\perp n} \= K^m{}_n \, , \qquad \qquad \Gamma^\perp_{\perp \mu} \= \Gamma^\mu_{\perp \perp} \= \Gamma^\perp_{\perp \perp} \= 0 \,.
\end{equation}
This is useful to decompose spacetime quantities into tangential and perpendicular components. For instance, let $V^\mu$ be a vector on the bulk manifold. A divergence of $V^\mu$ with respect to the bulk covariant derivative can be rewritten as
\begin{eqnarray}
	\nabla_\mu V^\mu = \bar{g}^{mn} \nabla_m V_n + \nabla_\perp V_\perp 
	= \bar{g}^{mn} \left(\mathcal{D}_m V_n - \Gamma^\perp_{mn} V_\perp \right) + \partial_\perp V_\perp 
	= \mathcal{D}_m V^m + K V_\perp + \partial_\perp V_\perp \, .
\end{eqnarray}
We can also derive the formula for $\delta K$ in \eqref{eqn: bdry cond 2} in the main text. The variation of the extrinsic curvature is given by \cite{compendium},
\begin{equation}\label{eqn: appendix Kmn}
	\delta K_{mn} \= -\frac{1}{2} \mathcal{D}_m h_{\perp n} -\frac{1}{2} \mathcal{D}_n h_{\perp m} - \frac{1}{2} K_{mn} h_{\perp \perp}  + \frac{1}{2}\partial_\perp h_{mn} \, . 
\end{equation}
Using this, we find that
\begin{equation}
\begin{split} 
	\delta K & \= \delta K_{mn} \bar{g}^{mn} - K_{mn} h^{mn} = -\mathcal{D}^m h_{\perp m} - \frac{1}{2} K h_{\perp \perp} + \frac{\bar{g}^{mn}}{2}\partial_\perp h_{mn} - K^{mn}h_{mn}   \\
	& \= -\mathcal{D}^m h_{\perp m} - \frac{1}{2} K h_{\perp \perp} + \frac{1}{2}\partial_\perp h^m{}_m \, ,
	\end{split}
\end{equation}
where in the last line we used that $K^{mn} = - \frac{1}{2}\partial_\perp \bar{g}^{mn}$.
 
It is also useful to see how various quantities transform under small bulk diffeomorphisms in Gaussian normal coordinates. Specifically, under $x^\mu \rightarrow x^\mu + \epsilon \, \xi^\mu(x)$,
\begin{eqnarray}
	\begin{cases}
		h_{mn} &\rightarrow \, h_{mn} + \mathcal{D}_m \xi_n + \mathcal{D}_n \xi_m + 2K_{mn} \xi_\perp \, , \\
		h_{\perp m} &\rightarrow \, h_{\perp m} + \partial_\perp \xi_m - 2 K_m{}^n\xi_n + \mathcal{D}_m \xi_\perp \, , \\
		h_{\perp \perp} &\rightarrow \, h_{\perp \perp} + 2\partial_\perp \xi_\perp \, .
	\end{cases}
\end{eqnarray}
Using \eqref{eqn: appendix Kmn}, one finds that 
\begin{equation}
	K_{mn} \,\rightarrow\, K_{mn} + \xi^p\mathcal{D}_p K_{mn} + K_{mp}\mathcal{D}_n \xi^p + K_{np} \mathcal{D}_m \xi^p - \mathcal{D}_{(m} \mathcal{D}_{n)} \xi_\perp + \left(\partial_\perp K_{mn}\right) \xi_\perp \, .
\end{equation}

\section{Derivation of the gauge-fixed solution for the flat boundary} \label{app:derivation}

In this appendix, we provide a detailed derivation of the solution \eqref{eqn: gauge-fixed sol kx!=0}. Recall that the gauge constraints $T_\mu(h_{\mu\nu})=0$ after imposing the boundary conditions \eqref{bdy_cond222} lead to the set of algebraic equations \eqref{eqn: gauge cond kx!=0}, 
\begin{equation}
\begin{dcases}
               \alpha_{mn} \= 0 \,, & \left( k^m k^n + k_x^2 \eta^{mn} \right)\beta_{mn} \= 0 \,, \\
               2ik^n \beta_{x n} - k_x \alpha_{x x} \= 0  \,, &  k_x  \beta^m{}_m -2ik^n \alpha_{x n} \= 0 \,, \\
               u^{(1)m} k_x \beta_{x m}  \= 0 \,,          &   i u^{(1)m} k^n \beta_{mn} - u^{(1)m} k_x \alpha_{x m}  \= 0 \,, \\
               u^{(2)m} k_x \beta_{x m}  \= 0 \,,          &   i u^{(2)m} k^n \beta_{mn} - u^{(2)m} k_x \alpha_{x m}  \= 0 \,.
        \end{dcases}
\end{equation}
By choosing the gauge $\alpha_{x\mu}=0$, these equations reduce to
\begin{equation}\label{appendix: dum1}
\begin{dcases}
               \alpha_{\mu \nu} \= 0 \,, &  \eta^{mn}\beta_{mn}   \= 0 \,, \\
               2ik^n \beta_{x n} \= 0  \,, & k^m k^n\beta_{mn}  \= 0 \,, \\
               u^{(1)m} k_x \beta_{x m}  \= 0 \,,          &   i u^{(1)m} k^n \beta_{mn}   \= 0 \,, \\
               u^{(2)m} k_x \beta_{x m}  \= 0 \,,          &   i u^{(2)m} k^n \beta_{mn}   \= 0 \,.
        \end{dcases}
\end{equation}
Since $k^m$, $u^{(1)m}$, and $u^{(2)m}$ span the tangent vector space on the timelike boundary, we can rewrite the boundary metric as
\begin{equation}\label{appendix: decompose eta}
	\eta^{mn} \= -\frac{k^m k^n}{k_x^2} + u^{(1)m}u^{(1)n} + u^{(2)m}u^{(2)n} \, .
\end{equation}
Hence, the condition $\eta^{mn}\beta_{mn}=0$ can be rewritten as
\begin{equation}
	-\frac{1}{k_x^2} k^m k^n\beta_{mn} + u^{(1)m}u^{(1)n}\beta_{mn} + u^{(2)m}u^{(2)n}\beta_{mn} \= 0 \, .
\end{equation}
Combining with the condition $k^mk^n \beta_{mn}=0$, we find
\begin{equation}\label{appendix: dum2}
	u^{(1)m}u^{(1)n}\beta_{mn} \= -  u^{(2)m}u^{(2)n}\beta_{mn} \, .
\end{equation}
 
Now we use \eqref{appendix: decompose eta} to express $\beta_{xm}$ as
\begin{equation}
	\beta_{xm} \= \delta^n_m \beta_{xn} \= - k_m \left(\frac{k^n \beta_{xn}}{k_x^2}\right) + u^{(1)}_m \left(u^{(1)n}\beta_{xn}\right) + u^{(2)}_m \left(u^{(2)n}\beta_{xn}\right) \, .
\end{equation}
The last three equations on the left column of \eqref{appendix: dum1} then imply that $\beta_{xm} = 0$. Similarly, we can express $\beta_{mn}$ as
\begin{equation}
\begin{split}
	\beta_{mn} \=  & \delta_m^p \delta_n^q \beta_{pq} 
	 \=  k_mk_n\left( \frac{k^pk^q\beta_{pq}}{k_x^4}\right) - 2k_{(m} u^{(1)}_{n)} \left(\frac{k^pu^{(1)q}\beta_{pq}}{k_x^2}\right) - 2k_{(m} u^{(2)}_{n)} \left(\frac{k^pu^{(2)q}\beta_{pq}}{k_x^2}\right)  \\
	& + u^{(1)}_{m}u^{(1)}_{n} \left(u^{(1)p}u^{(1)q}\beta_{pq}\right) + u^{(2)}_{m}u^{(2)}_{n} \left(u^{(2)p}u^{(2)q}\beta_{pq}\right) + 2u^{(1)}_{(m}u^{(2)}_{n)} \left(u^{(1)p}u^{(2)q}\beta_{pq}\right) \, .
	\end{split}
\end{equation}
The first three terms are zero according to the last three equations on the right column of \eqref{appendix: dum1}. The fourth and fifth terms are related through \eqref{appendix: dum2}. Finally, defining $\beta^{(+)} \equiv u^{(1)m}u^{(1)n}\beta_{mn}$ and $\beta^{(\times)} \equiv u^{(1)m}u^{(2)n}\beta_{mn}$, the remaining terms are 
\begin{equation}
	\beta_{mn} \= \left(u^{(1)}_{m}u^{(1)}_{n} - u^{(2)}_{m}u^{(2)}_{n} \right) \beta^{(+)} + 2u^{(1)}_{(m}u^{(2)}_{n)} \beta^{(\times)} \, .
\end{equation}
Plugging this into \eqref{eqn: flat non-compact sol 1}, we arrive at \eqref{eqn: gauge-fixed sol kx!=0} in the main text,
\begin{eqnarray}
	h_{\mu \nu}dx^\mu dx^\nu \= \left[\beta^{(+)} \left(u^{(1)}_n u^{(1)}_m - u^{(2)}_n u^{(2)}_m\right) + 2 \beta^{(\times)} u^{(1)}_n u^{(2)}_m \right]  \sin(k_x x) e^{i k_n x^n} dx^n dx^m \,.
\end{eqnarray}

\section{Kodama-Ishibashi formalism}\label{sec: Kodama-Ishibashi}

In this appendix, we review the Kodama-Ishibashi formalism \cite{Kodama:2000fa,Kodama:2003jz}. This formalism will allow us to deal with gravitational polarisations directly through gauge invariant quantities.
 
Let us consider a four-dimensional spacetime of the following type:
\begin{equation}
	ds^2 \= g_{ab}dy^a dy^b+ r(y)^2 d\sigma^2 \, , \label{eqn: metric decomposition}
\end{equation}
where $r(y)$ is an arbitrary function of $y^a$, $g_{ab}dy^ady^b$ is a two-dimensional Lorentzian manfiold, called an orbit space, and $d\sigma^2 = \sigma_{ij} dx^i dx^j$ is a metric of a two-dimensional maximally symmetric space. 
 
Any metric perturbation can be uniquely decomposed into vector ($V$) and scalar ($S$) perturbations,
\begin{equation}
    h_{\mu \nu} \= h_{\mu \nu}^{(V)} + h_{\mu \nu}^{(S)} \, ,
\end{equation}
where
\begin{equation}
\begin{cases}
    h_{a b}^{(V)}  \= 0 \, , \\
    h_{a i}^{(V)}  \= rf^V_a \,\mathbb{V}_i \, , \\
    h_{ij}^{(V)}  \= 2r^2H^V \,\mathbb{V}_{ij} \, ,  
    \end{cases}
    \qquad \qquad 
\begin{cases}
    h_{a b}^{(S)} \= f_{ab}\,\mathbb{S} \, , \\ 
    h_{a i}^{(S)} \= rf^S_a \,\mathbb{S}_i  \, , \\
    h_{ij}^{(S)} \= 2r^2 H^S \,\mathbb{S}_{ij} + 2r^2\gamma \, \sigma_{ij} \mathbb{S}\, .
 \end{cases}
\end{equation}
The coefficients of the perturbations, $f_{ab}$, $f_{a}^{V/S}$, $H^{V/S}$, and $\gamma$, are functions of the orbit coordinate $y^a$. The vector and scalar harmonic tensors $\mathbb{V}_i$, $\mathbb{V}_{ij}$, $\mathbb{S}$, $\mathbb{S}_i$, and $\mathbb{S}_{ij}$ are defined as followed. Let $\tilde{\mathcal{D}}_i$, $\tilde{\mathcal{D}}^2$, and $\mathcal{K}$ be covariant derivative, Laplacian, and unit curvature with respect to the metric $\sigma_{ij}$. The scalar harmonic function $\mathbb{S}$ is an eigenfunction of the scalar Laplacian with eigenvalue $-k_S^2$,
\begin{equation}
	\left(\Tilde{\mathcal{D}}^2+k_S^2\right) \mathbb{S} \= 0 \, .
\end{equation}
Given $\mathbb{S}$, one can construct a gradient $\mathbb{S}_i$ and an associated symmetric traceless rank-2 tensor $\mathbb{S}_{ij}$ as 
\begin{equation}\label{eqn: Kodama S_i and S_ij def}
	\mathbb{S}_i \,\equiv\, -\frac{1}{k_S}\Tilde{\mathcal{D}}_i\mathbb{S}\,, \qquad\qquad  \mathbb{S}_{ij} \, \equiv \, \frac{1}{k_S^2} \Tilde{\mathcal{D}}_i \Tilde{\mathcal{D}}_j \mathbb{S} + \frac{1}{2}\sigma_{ij}\mathbb{S}  \, ,
\end{equation}
which satisfy the following properties
\begin{equation}
\begin{cases}
	\left(\Tilde{\mathcal{D}}^2+k_S^2-\mathcal{K}\right) \mathbb{S}_i \= 0 \, , \\
	\Tilde{\mathcal{D}}^i \mathbb{S}_i \= k_S \mathbb{S} \,, 
	\end{cases}
	\qquad \qquad
\begin{cases}
	\left(\Tilde{\mathcal{D}}^2+k_S^2-4\mathcal{K}\right) \mathbb{S}_{ij} \= 0 \, , \\
	\Tilde{\mathcal{D}}^j\mathbb{S}_{ij} \= \frac{k_S^2-2\mathcal{K}}{2k_S}\mathbb{S}_i\,, \\
	\mathbb{S}^i{}_i \= 0 \, .
	\end{cases}
\end{equation}
The vector harmonic tensor $\mathbb{V}_i$ is a divergenceless eigenvector of the vector Laplacian with eigenvalue $-k_V^2$,
\begin{equation}
	\Tilde{\mathcal{D}}^i \mathbb{V}_i \= 0 \,, \qquad \left(\Tilde{\mathcal{D}}^2+k_V^2\right) \mathbb{V}_i \= 0 \, .
\end{equation}
Given $\mathbb{V}_i$, one can construct an associated symmetric traceless rank-2 tensor $\mathbb{V}_{ij}$ as
\begin{equation}\label{eqn: Kodama V_ij def}
	\mathbb{V}_{ij} \, \equiv \, -\frac{1}{2k_V} \left(\Tilde{\mathcal{D}}_i\mathbb{V}_j + \Tilde{\mathcal{D}}_j\mathbb{V}_i\right) \,,
\end{equation}
which satisfies the following properties
\begin{equation}
\begin{cases}
	\left(\Tilde{\mathcal{D}}^2+k_V^2-3 \mathcal{K}\right) \mathbb{V}_{ij} \= 0\, , \\
	\Tilde{\mathcal{D}}^j\mathbb{V}_{ij} \= \frac{k_V^2-\mathcal{K}}{2k_V}\mathbb{V}_i\,, \\ 
	\mathbb{V}^i{}_i \= 0 \, .	
	\end{cases}
\end{equation}
Explicit expressions for $\mathbb{S}$ and $\mathbb{V}_i$ in the case of a two-dimensional sphere are provided below. Note also that the $k_S = 0$, $2 \mathcal{K}$, and $k_V = \mathcal{K}$ modes are special and must be analysed separately \cite{Kodama:2000fa,Kodama:2003jz}. 
 
A diffeomorphism can be decomposed in a similar way, i.e., $ \xi_\mu = \xi_\mu^{(V)} + \xi_\mu^{(S)}$, with
\begin{equation}
\begin{cases}
    \xi_a^{(V)} \= 0 \, , \\ 
     \xi_i^{(V)} \= r\mathcal{L}^V \, \mathbb{V}_i \,, 
    \end{cases}
    \qquad \qquad
    \begin{cases}
    \xi_a^{(S)} \= \mathcal{T}_a \, \mathbb{S} \, , \\ 
    \xi_i^{(S)} \= r \mathcal{L}^S \, \mathbb{S}_i \, , 
    \end{cases}
    \label{eqn: Kodama diff decomposition}
\end{equation}
where $\mathcal{T}_a$ and $\mathcal{L}^{V/S}$ are functions of the orbit coordinates $y^a$. In terms of these, gauge transformation of the metric peturbation is given by
\begin{equation}
	\begin{cases}
		\delta f_a^V \= r \mathcal{D}_a \left(\frac{\mathcal{L}^V}{r}\right) \, , \\
		\delta H^V \= -\frac{k_V}{r}\mathcal{L}^V\, ,
	\end{cases}
	\qquad \qquad
	\begin{cases}
		\delta f_{ab} \= \mathcal{D}_a \mathcal{T}_b + \mathcal{D}_b \mathcal{T}_a \, , \\
		\delta f_a^S \= r \mathcal{D}_a \left(\frac{\mathcal{L}^S}{r}\right) - \frac{k_S}{r}\mathcal{T}_a \, , \\ 
		\delta H^S \= -\frac{k_S}{r}\mathcal{L}^S \, , \\ 
		\delta \gamma \=  \frac{k_S}{2 r}\mathcal{L}^S + \frac{\mathcal{D}_a r}{r}\mathcal{T}^a \,,
	\end{cases}
\end{equation}
where $\mathcal{D}_a$ is the covariant derivative with respect to $g_{ab}$.
By looking at the gauge transformation, gauge invariant quantities can be constructed as follows,
\begin{equation}\label{eqn: KI gauge invariant 1}
	\begin{cases}
		F_a & \equiv\, f_a^V + \frac{r}{k_V} \mathcal{D}_a H^V \, , \\
		F &\equiv\, \gamma + \frac{H^S}{2} + \frac{\mathcal{D}^a r}{r} X_a \, , \\
		F_{ab} & \equiv\, f_{ab} + \mathcal{D}_a X_b + \mathcal{D}_b X_a \, ,
	\end{cases}
\end{equation}
where $X_a \,\equiv\, \frac{r}{k_S}\left(f^S_a + \frac{r}{k_S}\mathcal{D}_a H^S\right)$.
 
By virtue of the Einstein field equation, $F_a$, $F$, and $F_{ab}$  can be expressed in terms of the so-called master fields $\Phi^V$ and $\Phi^S$ as follows:
\begin{equation}\label{eqn: KI gauge invariant 2}
	\begin{cases}
		F^a &=\, \frac{1}{r}\epsilon^{ab}\mathcal{D}_b \left(r \Phi^V\right) \, , \\
		F &=\, \frac{1}{8 r^2 }\Big[2\left(k_S^2 - 2\right)+4 r \mathcal{D}^ar\mathcal{D}_a\Big]\left(r \Phi^S\right)\, , \\
    		F_{ab} &=\, \Big[\mathcal{D}_a\mathcal{D}_b - \frac{1}{2} g_{ab}\Box\Big]\left(r \Phi^S\right) \, .
	\end{cases}
\end{equation}
The master fields $\Phi^{V/S}$ represent two polarisations of the gravitational fluctuation whose dynamics are governed by master equations
\begin{equation}
    \left( g^{ab}\mathcal{D}_a \mathcal{D}_b- V_{V/S}(r)\right)\Phi^{V/S} \= 0 \, ,
\end{equation}
where the effective potentials are given by
\begin{equation}
	V_V(r) \= \frac{k_V^2 + \mathcal{K}}{r^2} \, , \qquad \qquad V_S(r) \= \frac{k_S^2}{r^2} \, ,
\end{equation}
for vector and scalar perturbations, respectively. In this analysis, we have excluded the presence of a black hole, for simplicity. For formulae with a black hole or a non-zero $\Lambda$, see \cite{Kodama:2000fa,Kodama:2003jz}.

\subsection*{Two-dimensional sphere}\label{sec: two-sphere}

For a two-dimensional sphere, the metric $\sigma_{ij}$ and the unit curvature $\mathcal{K}$ are 
\begin{equation}\label{eqn: def two-sphere metric}
	d\sigma^2 \= d\theta^2 + \sin^2\theta d\phi^2 \, , \qquad \qquad \mathcal{K} \= 1 \, .
\end{equation}
The scalar/vector harmonics $\mathbb{S}$ and $\mathbb{V}_i$ can be constructed as follows. For a given angular momentum $l \in \mathbb{N}_0$, the scalar harmonic $\mathbb{S}$ is given by a spherical harmonic function,
\begin{equation}
	\mathbb{S} \= Y_{lm}(\theta,\phi) \, , \qquad k_S^2 \= l(l+1) \, ,
\end{equation}
where $m$ is an integer taking value between $-l \leq m \leq l$. The vector harmonic $\mathbb{V}_i$ is given by
\begin{equation}
	\mathbb{V}_i \= \left(\star d \mathbb{S}\right)_i \, , \qquad k_V^2 \= l(l+1)-1,
\end{equation}
where $d$ and $\star$ are exterior derivative and hodge dual operator associated to the metric \eqref{eqn: def two-sphere metric}. Their associated vector and symmetric traceless tensor harmonics, $\mathbb{S}_i$, $\mathbb{S}_{ij}$, and $\mathbb{V}_{ij}$ follow \eqref{eqn: Kodama S_i and S_ij def} and \eqref{eqn: Kodama V_ij def}.
 
When $l\=0$, only $\mathbb{S}$ is defined, and it becomes a constant function. When $l\=1$, $\mathbb{S}_{ij}$ and $\mathbb{V}_{ij}$ are not defined. Also, $\mathbb{V}_i$ becomes a Killing vector of the two sphere.

\bibliographystyle{JHEP}
\bibliography{bibliography}

\end{document}